\documentclass[aps,prd,12pt,nofootinbib,tightenlines,preprintnumbers,superscriptaddress,notitlepage]{revtex4-1}

\usepackage{graphicx}
\usepackage[utf8]{inputenc}
\usepackage{amssymb}
\usepackage{amsmath}
\usepackage{xcolor}
\usepackage{hyperref}
\usepackage{multirow}
\hypersetup{colorlinks, linkcolor = [rgb]{0,0.0,0.75}, citecolor = [rgb]{0,0.0,0.75}, urlcolor = [rgb]{0,0.0,0.75}}

\graphicspath{{./Figures/}}

\makeatletter
\g@addto@macro\bfseries{\boldmath}
\makeatother

\newcommand{\be}{\begin{equation}}
\newcommand{\ee}{\end{equation}}
\newcommand{\ba}{\begin{eqnarray}}
\newcommand{\ea}{\end{eqnarray}}
\newcommand{\la}{\label}

\newcommand{\<}{\langle}
\renewcommand{\>}{\rangle}

\newcommand{\ahlbl}{a_\mu^{\rm HLbL}}
\newcommand{\ahlblc}{a_\mu^{\rm HLbL,c}}
\newcommand{\chidof}{\chi^2/\text{dof}}

\newcommand{\metac}{M_{\eta_c}}

\begin{document}

\preprint{MITP-22-031}

\title{The charm-quark contribution to light-by-light scattering \\ in the muon $(g-2)$ from lattice QCD}

\author{En-Hung~Chao}
\affiliation{PRISMA$^+$ Cluster of Excellence \& Institut f\"ur Kernphysik,
Johannes Gutenberg-Universit\"at Mainz,
D-55099 Mainz, Germany}

\author{Renwick~J.~Hudspith}
\affiliation{PRISMA$^+$ Cluster of Excellence \& Institut f\"ur Kernphysik,
Johannes Gutenberg-Universit\"at Mainz,
D-55099 Mainz, Germany}

\author{Antoine~G\'erardin}
\affiliation{Aix Marseille Univ, Universit\'{e} de Toulon, CNRS, CPT, Marseille, France}

\author{Jeremy~R.~Green}
\affiliation{School of Mathematics and Hamilton Mathematics Institute, Trinity College, Dublin 2, Ireland}

\author{Harvey~B.~Meyer}
\affiliation{PRISMA$^+$ Cluster of Excellence \& Institut f\"ur Kernphysik,
Johannes Gutenberg-Universit\"at Mainz,
D-55099 Mainz, Germany}
\affiliation{Helmholtz~Institut~Mainz,
Staudingerweg 18, D-55128 Mainz, Germany}
\affiliation{GSI Helmholtzzentrum f\"ur Schwerionenforschung, Darmstadt, Germany}

\begin{abstract}
We compute the hadronic light-by-light scattering contribution to the
muon $g-2$ from the charm quark using lattice QCD. The calculation is performed
on ensembles generated with dynamical $(u,d,s)$ quarks
at the SU(3)$_{\rm f}$ symmetric point with degenerate pion and kaon masses of around 415\,MeV.
It includes the connected charm contribution, as well as the leading disconnected Wick contraction,
involving the correlation between a charm and a light-quark loop.
Cutoff effects turn out to be sizeable, which leads us to use lighter-than-physical charm
masses, to employ a broad range of lattice spacings reaching down to 0.039\,fm and
to perform a combined charm-mass and continuum extrapolation.
We use the $\eta_c$ meson to define the physical charm-mass point and 
obtain a final value of $\ahlblc = (2.8\pm 0.5) \times 10^{-11}$,
whose uncertainty is dominated by the systematics of the extrapolation.
Our result is consistent with the estimate based on a simple charm-quark loop,
whilst being free of any perturbative scheme dependence on the charm mass.
The mixed charm--light disconnected contraction contributes a small negative amount to the final value.
\end{abstract} 

\date{\today}

\maketitle

\section{Introduction}

The anomalous magnetic moment of the muon, $a_\mu \equiv (g-2)_\mu/2$,
is one of the most precisely measured quantities in fundamental
physics. Currently, the experimental world
average~\cite{Muong-2:2021ojo,Bennett:2006fi} and the theoretical
evaluation of the 2020 White Paper (WP)~\cite{Aoyama:2020ynm} based on
the Standard Model (SM) of particle physics are in tension at the
$4.2\sigma$ level. The theory uncertainties are entirely dominated by
the hadronic contributions. Surprisingly, a lattice-QCD based
calculation~\cite{Borsanyi:2020mff} of the leading hadronic
contribution finds a larger value than the dispersion-theory based
estimate of the WP, which would bring the overall theory prediction
into far better agreement with the experimental value of $a_\mu$. Thus
it will be vital to resolve the tension between the different
determinations of the leading hadronic contribution in order to
strengthen the unique test of the SM offered by the anomalous magnetic
moment of the muon.

A subleading hadronic contribution to $a_\mu$, the hadronic
light-by-light (HLbL) contribution, also contributes sizeably to
the error budget of the SM prediction. The HLbL contribution is
significantly more complex to evaluate than the leading hadronic
contribution; however, because it is suppressed by an additional power
of the fine-structure constant $\alpha$, it only needs to be
determined at the ten percent level.  The HLbL contribution, too,
has been evaluated using either dispersive
methods~\cite{Aoyama:2020ynm} or lattice
QCD~\cite{Blum:2019ugy,Chao:2021tvp}.  In this case, good agreement is
found among the three evaluations within the quoted uncertainties.

One missing ingredient in the otherwise complete HLbL calculation
of~\cite{Chao:2021tvp} is the contribution of the charm quark.
The present paper addresses this missing contribution.
Since the charm quark is much heavier than the muon, on general grounds~\cite{Beresetskii:1956,Cowland:1958}
one expects this contribution to be in a regime where it is
roughly proportional to $m_\mu^2/m_{c}^2$.
In phenomenological estimates, it has been evaluated using the
prediction based on Quantum Electrodynamics (QED), amended for the appropriate charge and colour
factors.  We quote the value and uncertainty from the 2020 White
Paper~\cite{Aoyama:2020ynm},
\be\la{eq:amucWP}
a_\mu^{\rm HLbL,c}({\rm WP}) = (3 \pm 1) \times 10^{-11}.
\ee
The main goal of this paper is thus to test the prediction (\ref{eq:amucWP}) using lattice QCD,
in case a qualitative effect might have been missed.
Certainly, this contribution is small compared to the overall uncertainty $43\times 10^{-11}$ of the WP prediction
for $a_\mu$, however the other uncertainties are also expected to shrink, especially if the issues
in the leading hadronic contribution can be resolved.

Our second motivation for addressing the charm HLbL contribution from
first principles is to answer the qualitative question whether
approximating this contribution via a simple quark loop is
adequate. In lattice QCD, the calculation involves computing charm
propagators on an ensemble of non-perturbative background SU(3) gauge
fields. If the simple quark-loop picture is approximately correct, the
details of this gauge field should not matter much, and the charm
propagators can be replaced by free Dirac propagators.  In this case,
the sensitivity to the sea quarks enters (at the earliest) at quadratic order in $\alpha_s(m_c)$,
the strong coupling constant at the scale of the charm mass.
It is largely for this reason that we will focus on the SU(3)$_{\rm f}$-symmetric mass point with
$m_\pi=m_K\simeq 415$\,MeV, enabling us to reach sufficiently fine lattices at a moderate computational cost.

A further aspect of the quark-loop picture is that the various disconnected diagrams entering
the HLbL amplitude are expected to be small.
In contrast, if the $\eta_c $ pole exchange or $D$ meson loops played a sizeable role in the
charm-quark contribution, the leading disconnected
charm contribution, consisting of a charm loop and a light-quark loop,
each attached to two electromagnetic currents, would be sizeable
(in analogy to the analyses in~\cite{Bijnens:2016hgx} and appendix~A of Ref.~\!\cite{Chao:2021tvp} for the three-flavour case).
We recall that for the light quarks, individual mesons, especially the
pseudoscalars $\pi^0,\eta,\eta'$, contribute substantially to
$a_\mu^{\rm HLbL}$, even at the aforementioned SU(3)$_{\rm
  f}$-symmetric point~\cite{Chao:2020kwq}. 
In lattice QCD, we can quantitatively test the relevance of the disconnected contributions.

This paper is organized as follows. We describe our lattice setup, the
tuning of the charm quark mass and our specific representation of
$\ahlbl$ in section~\ref{sec:setup}.  Section~\ref{sec:theo} provides
some basic theory expectations concerning the connected and leading
disconnected contributions involving a charm quark.
Section~\ref{sec:conn} presents our lattice results on the connected
contribution for a sequence of increasing charm-quark masses, and
section~\ref{sec:disc} contains our results at the target charm mass
for the leading topology of disconnected diagrams.  We provide our
final result and conclude in section~\ref{sec:concl}.
Appendix~\ref{sec:lloop_app} describes a test of our methods at a
heavy quark mass in lattice QED, while
appendix~\ref{sec:data_tables_conn} contains tables of results for the
connected charm contribution on individual ensembles and appendix
\ref{sec:fit_results} a representative set of fit results.

\section{Lattice setup \la{sec:setup}}

We have performed lattice-QCD calculations on gauge ensembles provided by the
Coordinated Lattice Simulations (CLS) initiative~\cite{Bruno:2014jqa},
which have been generated using three flavours of non-perturbatively
O($a$)-improved Wilson-clover fermions and with the tree-level-improved L\"uscher-Weisz  gauge action.
As in Ref.~\!\cite{Chao:2020kwq}, where we computed the $(u,d,s)$ quark contribution,
we consider only ensembles realizing exact SU$(3)_{\rm f}$-symmetry.
On these ensembles, the mass of
the octet of light pseudoscalar mesons is approximately 415~MeV. The parameters of these
ensembles, which correspond to six different values of the lattice spacing,
are summarized in Table~\ref{tab:ensembles}.

\begin{table}
  \centering
  \begin{tabular}{c|cccc|lc}
  \toprule
Label & $\beta$ & $\kappa$ & $L^3\times L_T$ & Temporal B.Cs & $a$ (fm) & $m_{\pi,K,\eta}$ (MeV) \\ 
\hline
	  A653 & 3.34 & 0.1365716  & $24^3\times 48$  & periodic &  0.09930(\textbf{122})     & 413(5)${}^\ast$ \\
H101 & 3.40 & 0.13675962 & $32^3\times 96$  & open     &  0.08636(98)(40) & 418(5) \\
B450 & 3.46 & 0.13689    & $32^3\times 64$  & periodic &  0.07634(92)(31) & 417(5) \\
N202 & 3.55 & 0.137000   & $48^3\times 128$ & open     &  0.06426(74)(17) & 412(5) \\
N300 & 3.70 & 0.137000   & $48^3\times 128$ & open     &  0.04981(56)(10) & 421(5) \\
	  J500 & 3.85 & 0.136852   & $64^3\times 192$ & open     &  0.03910(\textbf{46})       & 413(5)${}^\ast$ \\
\botrule
  \end{tabular}
  \caption{The SU$(3)_{\rm f}$-symmetric  ensembles used in this work.
    Each ensemble is parametrized by the gauge coupling parameter $\beta\equiv 6/g_0^2$,
    the $(u,d,s)$-quark hopping parameter $\kappa$, the lattice size, and the temporal
    boundary condition. The lattice spacings $a$ were determined in Ref.~\cite{Bruno:2016plf}, apart from A653 and J500, where the lattice spacings were estimated from the ratio of the Wilson flow parameter $t_0$; the errors on the lattice spacing for these two ensembles (in bold) are simply estimated by scaling of the total error of the neighboring lattice spacings. Their pion masses (marked with asterisk) have been measured independently for this work.}
  \label{tab:ensembles}
\end{table}

\subsection{Calibrating the charm mass and current}

The connected contribution to $\ahlbl$ and the two-point correlation
function of $\bar c \gamma_5 c$ were computed on all ensembles of
Table~\ref{tab:ensembles} for several (5 or 7) values of the charm-quark
bare subtracted  mass $am_{\rm c}=(\kappa_c^{-1}-\kappa_{\rm crit}^{-1})/2$, with values of $\kappa_c$ chosen to interpolate
between the physical strange and charm hopping parameters.
A determination of the latter is available from Ref.~\!\cite{gerardin:2019rua},
obtained by tuning the $D_s$ meson mass to its physical value.
For the (dominant) connected contribution however, we choose the
physical charm-mass point as the one defined by the physical value of
the $\eta_c$ meson mass. When we determine the $\eta_c$ mass, we do not include the disconnected diagram
in the two-point function of the charm pseudoscalar density.
This procedure corresponds to using the
operator $\bar c' \gamma_5 c$ ($am_{\rm c^\prime} = am_{\rm c}$), where the
degenerate quark flavours $c$ and $c'$ are both treated at the
partially-quenched level. It should be noted that
the tuning of Ref.~\!\cite{gerardin:2019rua} by the $D_s$ yields
a heavier-than-physical $\eta_c$ meson at our $\text{SU}(3)_\text{f}$-symmetric point.
This comes from the quark masses at the latter point being lighter than the physical strange quark,
and a $D_s$-tuning \emph{de facto} absorbs this effect into the charm-quark mass~\cite{Hudspith:2021iqu}.

The reason for using lighter-than-physical charm quark masses is that
we expect discretisation effects to become more and more significant
when the charm mass increases. For a rough estimate of the typical
size of discretisation effects, \cite{Hudspith:2021iqu} found that the
effective speed of light (as defined by the dispersion relation of a
meson)
for physical-mass charm quarks at worst deviates from unity by 20\% in our setup.

The finite renormalisation factor $Z_V^c(g_0,am_{\rm c})$ for the
local charm current $\bar c \gamma_\mu c $ was determined by requiring
the corresponding charge of the ground-state meson created when $\bar
c' \gamma_5 c$ acts on the vacuum to be unity.
The meson correlators were computed using $Z_2 \times Z_2$ stochastic wall sources \cite{Foster:1998vw, Boyle:2008rh}.
The quark-mass dependence of $Z_V^c(g_0,am_{\rm c})$
is quite strong, especially at coarse lattice spacings. Since
this factor enters to the third power into our final result, we determine
it directly for every one of the bare quark-mass values. This is the
same procedure that was implemented for the charm renormalisation in
\cite{gerardin:2019rua}.\footnote{In that paper, the $Z_V^c$ values
quoted in its appendix~A  are erroneously 
described as stemming from the charm number of the $D_s$ meson.}

\subsection{Computing the charm contribution to $a_\mu^{\rm HLbL}$}

We apply the formalism described and used in~\cite{Chao:2020kwq,Chao:2021tvp} and therefore only recall the main aspects.
The starting point of our calculation is the master formula\footnote{See however the text below Eq.\ (\ref{eq:emcurr})
for references to the precise formulae used in the present calculation.}
\begin{equation}
a_\mu^{\rm HLbL} = \int_0^\infty d|y|\; f(|y|),\quad f(|y|) = \frac{m_\mu e^6}{3} 2\pi^2  |y|^3
\int_x \; \bar{\cal L}_{[\rho,\sigma];\mu\nu\lambda}(x,y)\;i\widehat\Pi_{\rho;\mu\nu\lambda\sigma}(x,y).
\end{equation}
Here $e^2/(4\pi)=\alpha_{\rm QED}$ is the fine-structure constant and $m_\mu$ the muon mass.
The QED kernel $\bar{\cal L}$ has been computed in the continuum~\cite{Asmussen:2016lse}
and represents the contributions of the photon and muon propagators and vertices
in the diagrams of Fig.~\ref{fig:Wick}. There is a lot of freedom to alter the kernel without changing $a_\mu^{\rm HLbL}$ in the continuum and in infinite volume. Specifically, we use the kernel $\bar{\cal L}^{\Lambda}$ defined in~\cite{Chao:2020kwq} with $\Lambda=0.40$ throughout.
The tensor $i\widehat\Pi$ is a Euclidean hadronic four-point function with one of its vertices weighted linearly in one of its coordinates,
\begin{equation}\label{eq:pihat}
\begin{aligned}
  i\widehat \Pi_{\rho;\mu\nu\lambda\sigma}( x, y)  &= -\int_z  z_\rho\,
  \Big\<\,j_\mu(x)\,j_\nu(y)\,j_\sigma(z)\, j_\lambda(0)\Big\>_{\rm QCD}.
\end{aligned}
\end{equation}
The field $j_\mu(x)$ appearing above is the hadronic component of the electromagnetic current,
\begin{equation}\la{eq:emcurr}
  j_\mu(x) = \sum_f {\cal Q}_f \;(\bar q_f \gamma_{\mu} q_f)(x).
\end{equation}

Here we focus on the contributions involving the charm current, $\bar
c\gamma_\mu c$.  The QCD  four-point function
 receives contributions from five classes of Wick
contractions.  First, we will focus on the fully-connected charm
contribution, which involves four charm currents; for this contribution,
we  apply Eq.\ (7) of Ref.~\!\cite{Chao:2021tvp} with the flavour index set to charm, $j:=c$.
Second, we will consider the disconnected contributions involving two quark loops,
each of which containing two vector vertices, with either one or both
loops consisting of charm propagators. Here we apply
Eq.\ (11) of Ref.~\!\cite{Chao:2021tvp} with the flavour indices $i,j$ running
over $\{u,d,s,c\}$ under the constraint that at least one of them take the value $c$. 
The connected and (leading) disconnected contributions are illustrated in Fig.~\ref{fig:Wick}.

\begin{figure}
  \centerline{\includegraphics[width=0.25\textwidth]{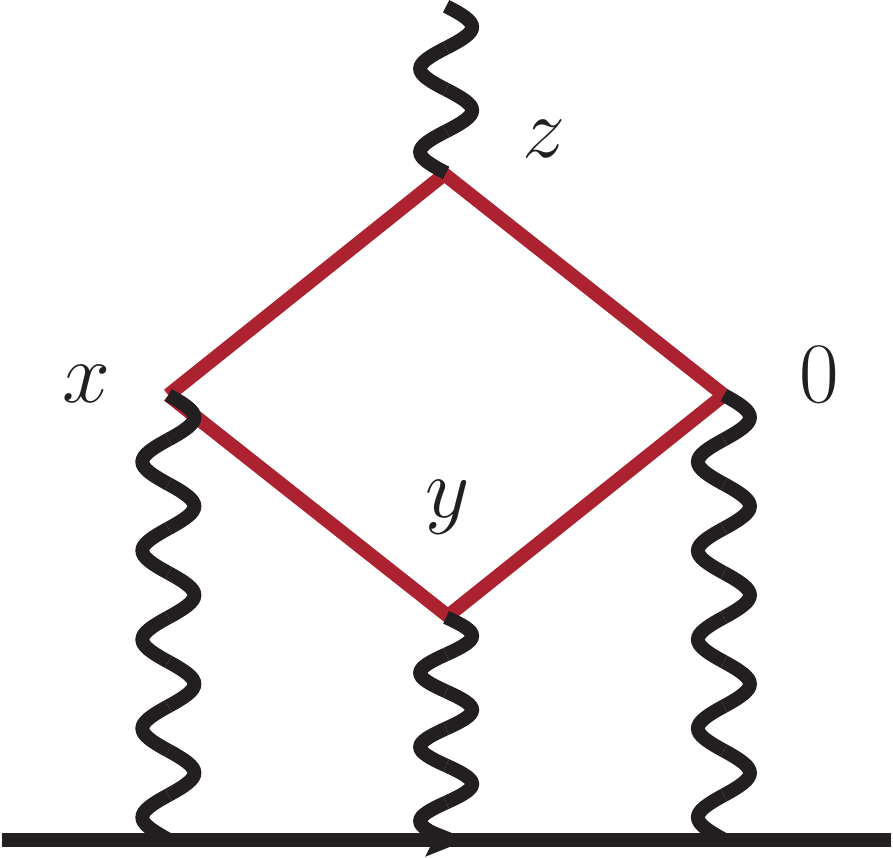}
    ~~~~~~~~~~~~~~~~~~~~~\includegraphics[width=0.25\textwidth]{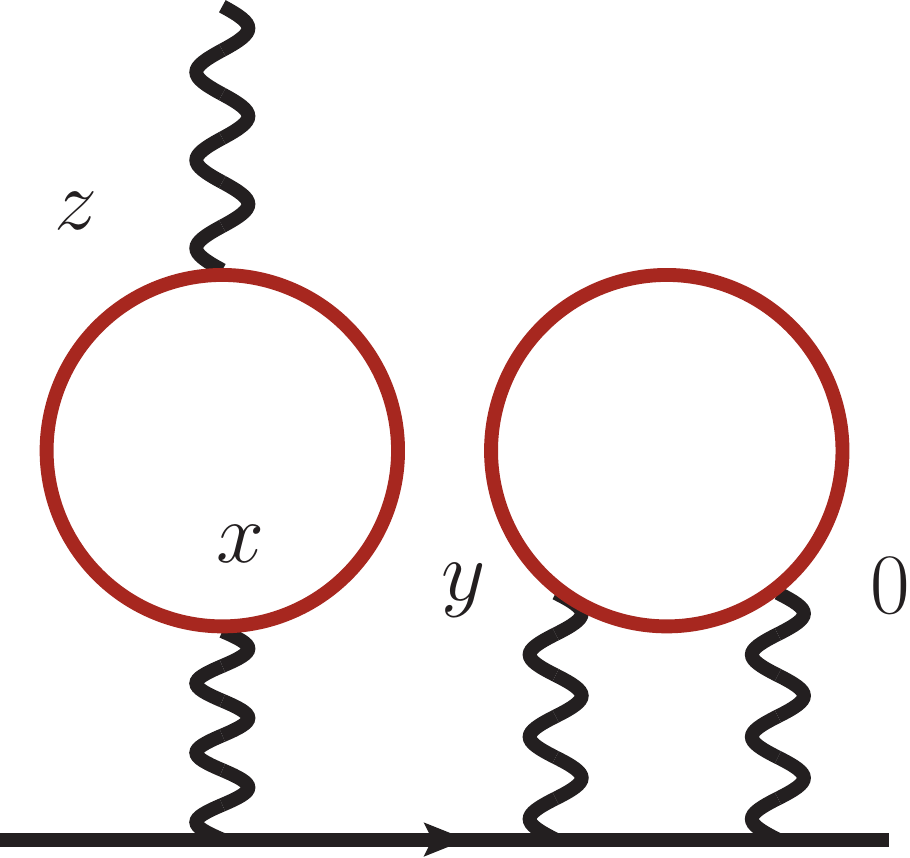}}
\caption{\la{fig:Wick} The fully connected charm contribution (left) and the (2+2) Wick contraction (right, with
at least one loop corresponding to a charm quark) are the two Wick contractions computed in this work.}
\end{figure}

\section{Theory expectations\la{sec:theo}}

The simplest prediction for the light-by-light contribution of a heavy `charm' quark to $10^{11} a_\mu$
relies on the analytic QED result originally applied to the $\tau$ lepton loop~\cite{Laporta:1992pa,Kuhn:2003pu}.
Taking into account the colour factor $N_c=3$ and the charge factor $(2/3)^4$, it is given by the function
\be\la{eq:leploop_analytic}
h(m_Q) =  5.10382  \frac{1}{\hat m_Q^2} +
\Big(-0.176225  -0.0567645  \log(\hat m_Q^2)  -0.00459931 \log^2(\hat m_Q^2) \Big) \frac{1}{\hat m_Q^4},
\ee
with $\hat m_Q$ is the heavy-quark mass in GeV. Already by $m_Q=0.75$\,GeV, the O$(m_Q^{-4})$ terms only
represent a reduction of the leading term by five percent. These terms certainly represent a small correction
for $m_Q$ around the physical charm mass. Here we have dropped known higher-order terms in $1/m_Q$.
We will take the function $h(m_Q)$  as a baseline for comparison with our lattice results for the fully connected charm contribution.

For the (2+2) disconnected contribution involving one charm and one light-quark loop,
it is less straightforward to make a `baseline' prediction.
The scalar-QED prediction for the contribution to $\ahlbl$ of the
$D^\pm$ meson loop is $-0.33\times 10^{-11}$~\cite{Kuhn:2003pu}, to be roughly 
doubled in order to include the $D_s$ loop. Taking into account the charge
factor of $2\cdot3\cdot {\cal Q}_c^2({\cal Q}_u^2+{\cal Q}_d^2+{\cal Q}_s^2)=\frac{144}{81}$
relevant for the charm--light (2+2) contribution
(see~\cite{Chao:2021tvp}, appendix~A\footnote{The analysis of this reference can be applied here due to
the $u,d,s$ quarks being degenerate and the charm quark being quenched.}), one arrives at the prediction of
$a_\mu^{2+2:lc}= -0.58\times 10^{-11}$ when treating the $D^+,
D^0,D_s$ meson loops within scalar QED.\footnote{Note that the $D^0$ loop
contributes to $a_\mu^{2+2:lc}$, even though at the scalar-QED level it does not contribute to $\ahlbl$,
due to it cancelling between different topologies.}
The absolute value of this prediction
is surely an overestimate, given that electromagnetic form factors of
the $D$ mesons should suppress this prediction substantially: in the
case of the pion loop, the suppression factor is almost three, and for
the kaon almost ten~\cite{Aoyama:2020ynm,Kuhn:2003pu}. All in all, these considerations
finally lead us to expect an order of magnitude of $(-0.3\pm0.3)\times 10^{-11}$.

In addition to the short-distance effect estimated above,
the charm--light disconnected diagrams also involve a longer-distance contribution, whose size it is useful
to estimate by theory arguments, given the difficulty of measuring the correlation function in the infrared.
The intuitive idea is that the heavy-quark loop shrinks almost to a point in coordinate-space\footnote{This picture holds when
the vertices of the light-quark loop are at a distance much greater than $(2m_c)^{-1}$ from the charm loop.},
acting effectively like a local gauge-invariant gluonic operator from the point of view of the `low-energy effective theory',
which is QCD with $(u,d,s)$ quarks. This picture can be formalized by writing an effective Lagrangian for the effective
coupling induced between two photons and gluonic fields, much as in the classic work of Euler and Heisenberg~\cite{Heisenberg:1936nmg}.
This effective Lagrangian ${\cal L}^{(c)}_{2\gamma2g}$ has been calculated long ago~\cite{Novikov:1979va};
each term of the Lagrangian contains two photonic and two gluonic field strength tensors.
From here, one infers the operator equation 
\be\la{eq:L2gam2g}
{\cal Q}_c^2\;{\rm T}\{ (\bar c\gamma^\mu c)(x)\; (\bar c\gamma^\mu c)(y)\} 
= \frac{1}{e^2}\frac{\delta^2}{\delta A_\mu(x) \delta A_\nu(y)} \int d^4w\; {\cal L}^{(c)}_{2\gamma2g}(w),
\ee
$A_\mu$ being the photon field,
which shows that the charm loop acts at low energies like a set of gluonic operators such as $\alpha_s G_{\mu\nu}^a G_{\mu\nu}^a$
or $\alpha_s G_{\mu\nu}^a \tilde G_{\mu\nu}^a$.
The main observation is that, on dimensional grounds, the effective Lagrangian is overall multiplied by a $1/m_c^4$ factor,
indicating a strong suppression.

The argument above shows that a light flavour-singlet meson such as
the scalar $f_0$ or the pseudoscalar $\eta'$ can propagate between the
charm loop and the light-quark loop, albeit with a very suppressed
coupling to the charm loop.  To get an estimate of this contribution,
which is long-range in comparison to the length-scale $(2m_c)^{-1}$,
we use Eq.\ (\ref{eq:L2gam2g}) to find out roughly how much the charm
part of the electromagnetic current by itself contributes to the
$\eta'$ transition form factor (TFF). Note that this contribution is
independent of the photon virtualities, as long as these are small.
Using the estimate $\<0|\alpha_s G \tilde G | \eta'\>\approx 0.5\,{\rm
  GeV}^3$ based on Ref.~\cite{Novikov:1979uy}, while the TFF
normalisation amounts to $|{\cal F}_{\eta'\gamma\gamma}(0,0)| \simeq
0.34\, {\rm GeV}^{-1}$ (see for instance~\cite{Nyffeler:2016gnb}), we
obtain a contribution of about $8\times 10^{-4}\,{\rm GeV}^{-1}$ to
${\cal F}_{\eta'\gamma\gamma}$ from the charm current.  Since the
$\eta'$ exchange contributes about $14.5\times 10^{-11}$ to
$a_\mu^{\rm HLbL}$~\cite{Aoyama:2020ynm}, proportionally to its TFF at
each end of $\eta'$ propagator, we arrive at the order-of-magnitude
estimate of $0.01\times 10^{-11}$ for the contribution to $a_\mu^{\rm
  HLbL}$ of the $\eta'$ in the (2+2) charm--light diagrams.  Even with
a potential logarithmic enhancement~\cite{Knecht:2001qg}, this is much
smaller than our final uncertainty and cannot presently be resolved in
our lattice calculations.  

In addition to the Wick-contraction topologies considered above, the
(3+1) topology with the single-current loop consisting of a charm
propagator deserves some attention, since this contribution is neither
SU(3)$_{\rm f}$ nor $1/N_c$ suppressed relative to the (2+2) topology,
$N_c$ being the number of colours.  In perturbation theory, the (3+1)
contribution starts at O($\alpha_s^3$) rather than at O($\alpha_s^2$),
while involving the same minimal number of charm propagators.
Furthermore, the quark-charge and multiplicity factors numerically
suppresses this contribution by a relative factor of three\footnote{Within the scalar QED framework,
the two topologies by themselves contain
equal and opposite contributions from the $D$ meson loops, since they cancel in
$ \< \bar c \gamma_\mu c\; \bar u \gamma_\nu u\;\bar u \gamma_\rho u\; \sum_{f=u,d,s,c} \bar q_f \gamma_\lambda q_f \>$,
given that the charge of $D$ mesons is zero under the total quark number current $\sum_{f=u,d,s,c} \bar q_f \gamma_\lambda q_f $.}
since it is weighted by $4\cdot({\cal Q}_u^3+{\cal Q}_d^3+{\cal Q}_s^3){\cal
  Q}_c = 48/81$, while the charm--light (2+2) diagrams are weighted by
$144/81$, as noted above. A factor of three suppression relative to the
(2+2) charm--light contribution is thus expected.

\section{Lattice results for the connected contribution\la{sec:conn}}

As a way of validating our computational methods, 
 our analysis has been guided by a
 lepton-loop calculation, much like in Ref.~\!\cite{Chao:2020kwq}:
in Appendix~\ref{sec:lloop_app} we
investigate the applicability of our QED-kernel implementation at
particularly heavy scales by comparing the lepton-loop contribution to $a_\mu^{\rm LbL}$
to the known analytical expression~\cite{Laporta:1992pa}. While
the agreement is acceptable at fairly heavy lepton mass, the study
suggests that cut-off effects will be significant and working at
unphysically-light charm mass might allow for a better handle on these
effects. The physical charm mass will therefore be approached via a simultaneous extrapolation
in the quark mass and in the lattice spacing.

\subsection{Results at individual quark masses}

For the connected part of $\ahlblc$, we have performed computations
with the vector current connected to the external on-shell photon (the
$z$-vertex in Eq.~(\ref{eq:pihat})) being either symmetrised-conserved\footnote{A definition
of the local and the symmetrised-conserved current can be found for instance in Ref.\ \cite{Gerardin:2018kpy}.}
or local, while the rest of the currents are kept local.
For each ensemble, we have tuned $\kappa_c$ to get five to seven different
$\eta_c$-masses, ranging from around 1.3 to 2.6 GeV. In order to
better control rotational-symmetry breaking effects (and keep the
higher-order lattice artifact coefficients the same) we will only use
$f(|y|)$ along the lattice direction (1,1,1,1) for all ensembles.

\begin{figure}[h!]
\includegraphics[scale=0.275]{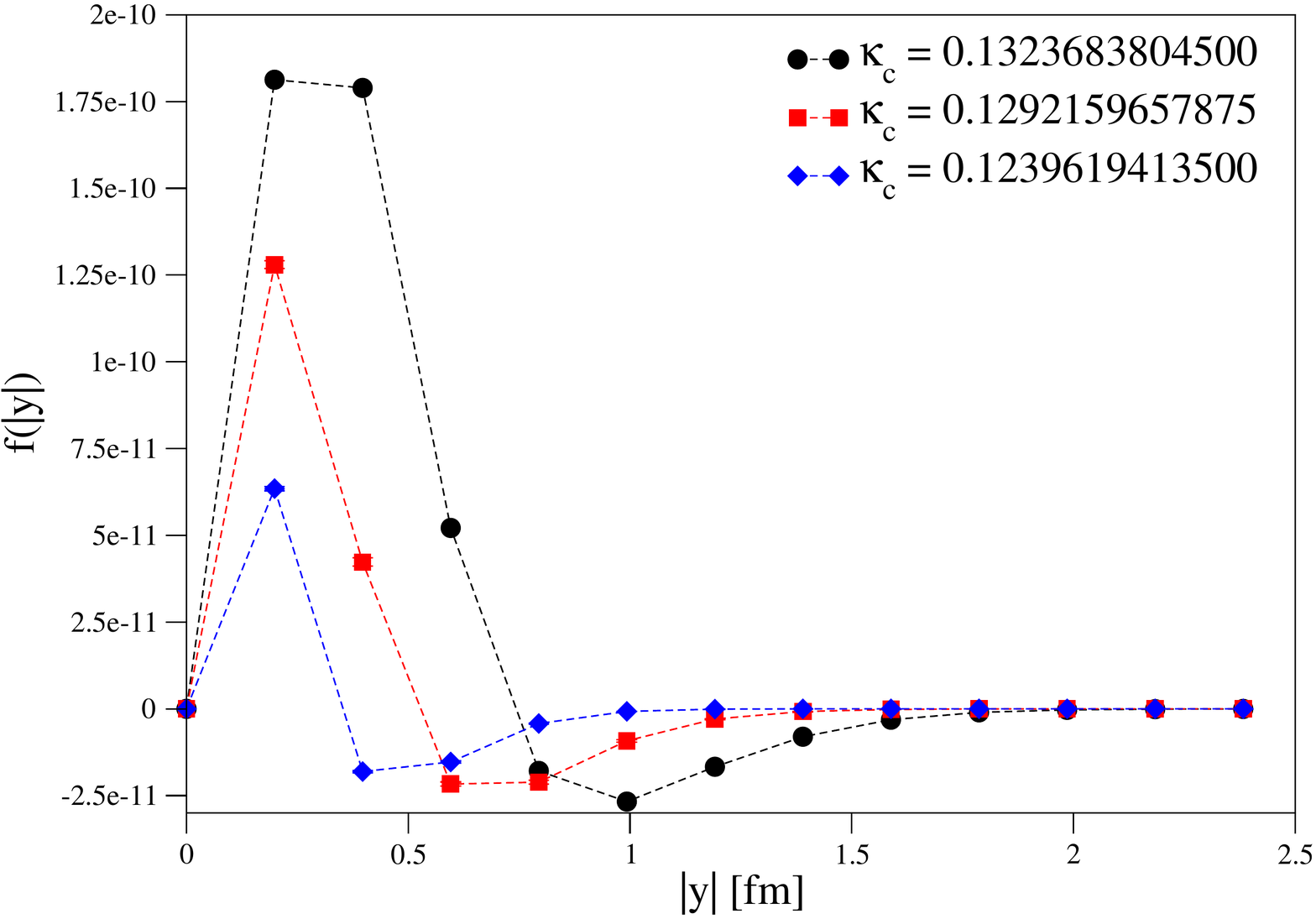}
\hspace{-8pt}
\includegraphics[scale=0.275]{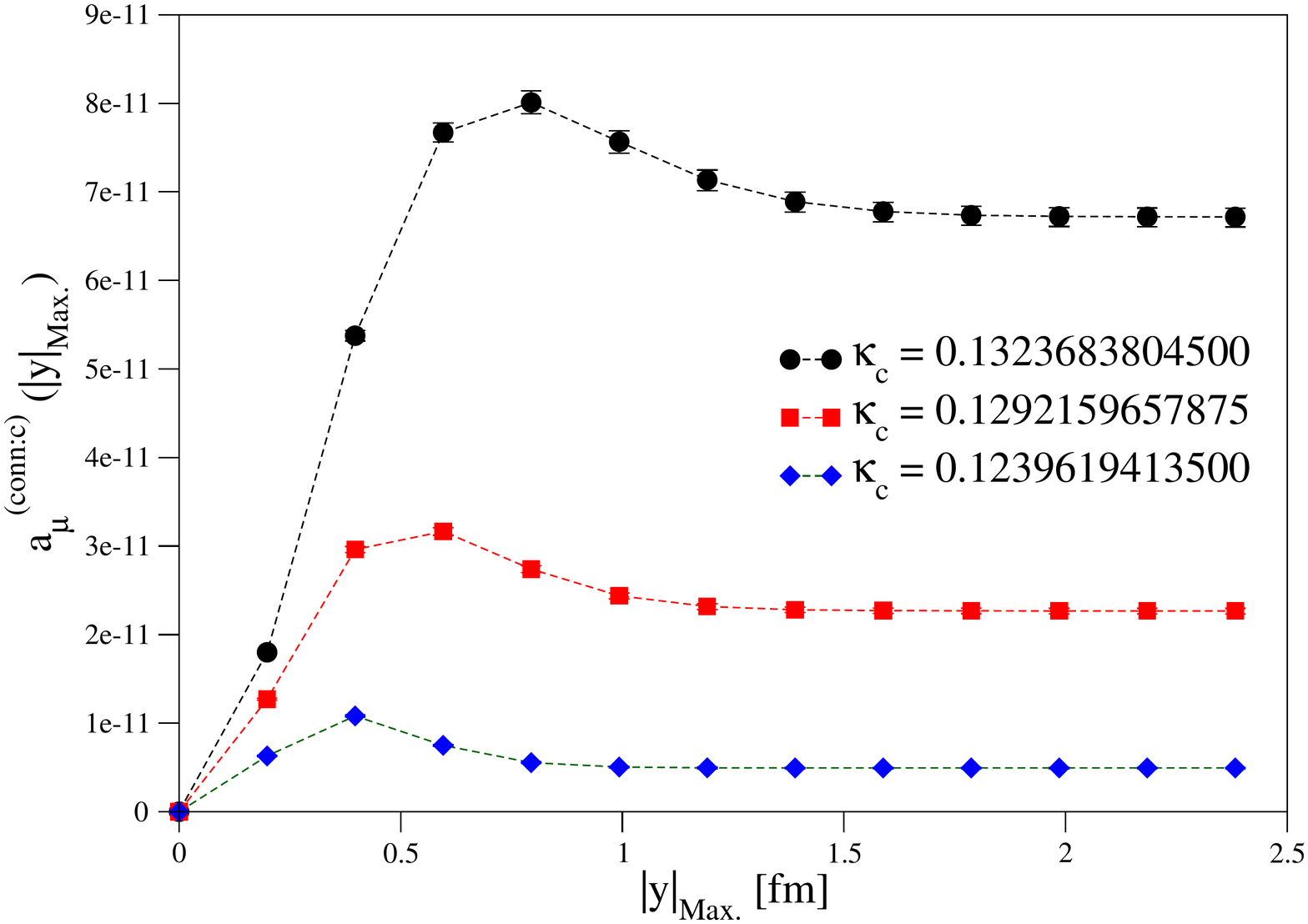}
\caption{The local-conserved integrands (left) and partially-integrated results (right) for a selection of $\kappa_c$s (heaviest, middle, and lightest $a\metac$) for our coarsest ensemble A653. Dashed lines are to guide the eye.}\label{fig:ABOX_cont}
\end{figure}

\begin{figure}[h!]
\includegraphics[scale=0.275]{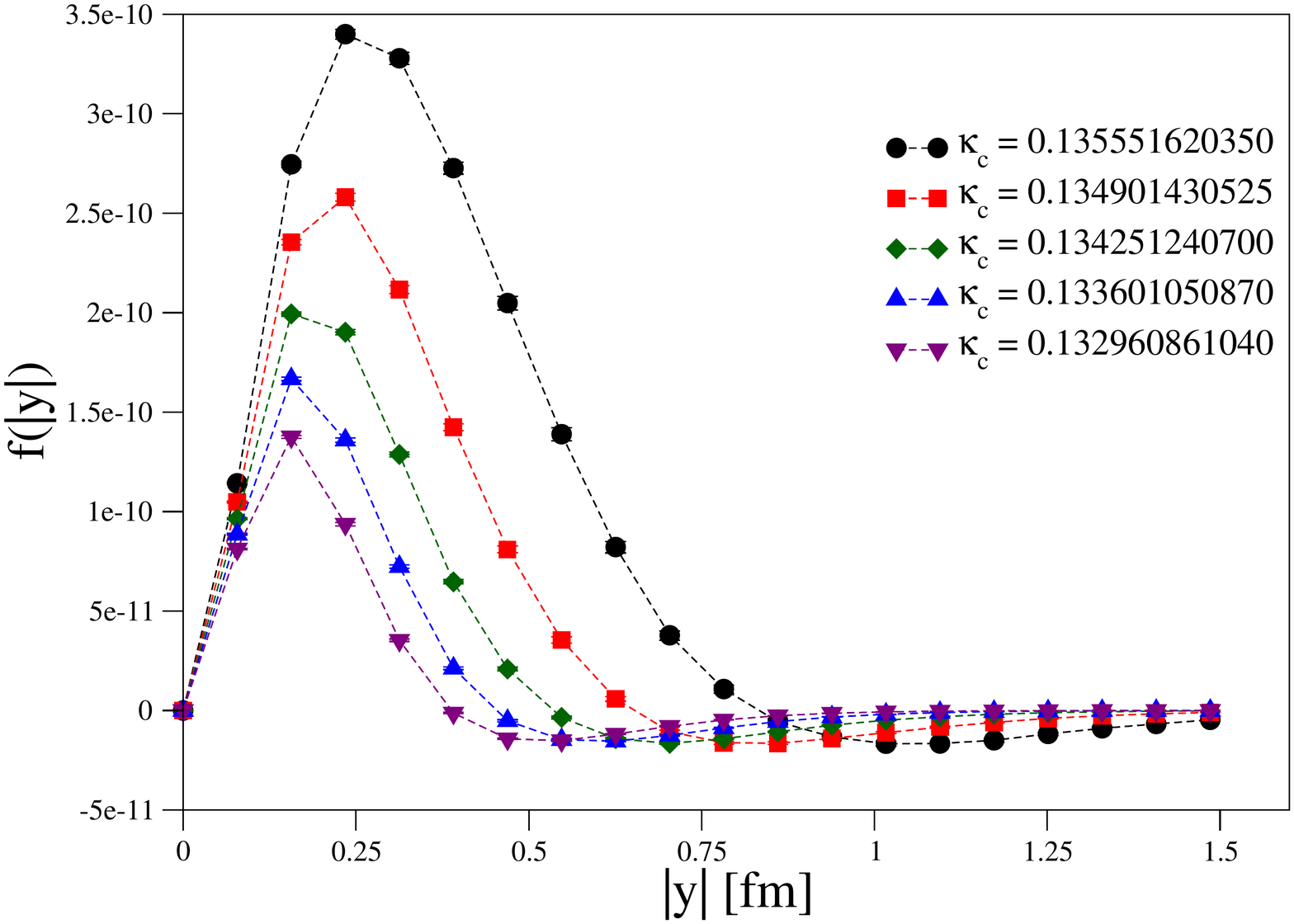}
\hspace{-8pt}
\includegraphics[scale=0.275]{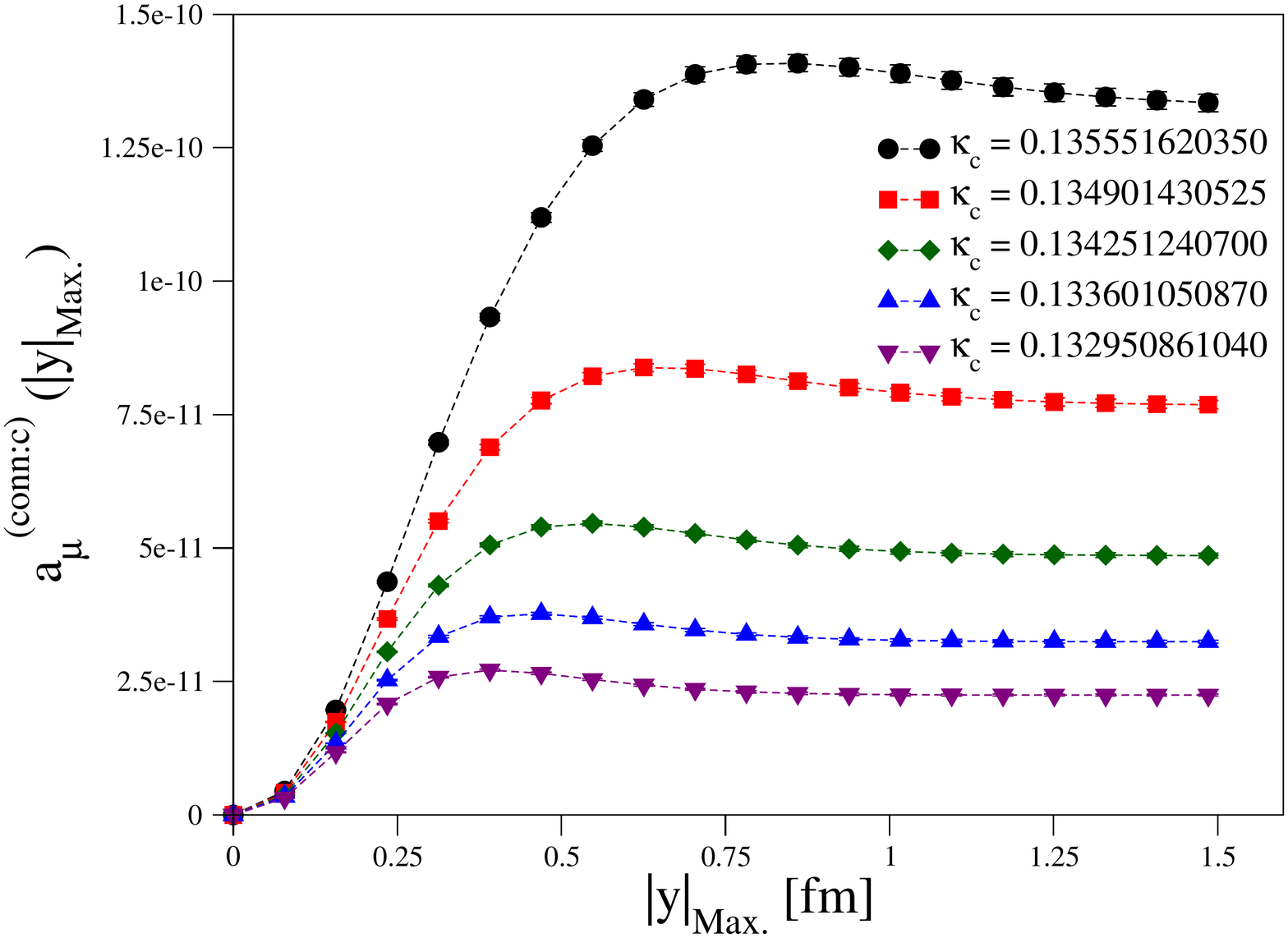}
\caption{The local-conserved integrands (left) and partially-integrated results (right) for all of our $\kappa_c$s, from our finest ensemble J500.}\label{fig:JBOX_cont}
\end{figure}

Fig.~\ref{fig:ABOX_cont} shows an example of our data for the A653
ensemble. The integrand is steeply-peaked at short distances and
becomes more so at heavier quark masses (smaller $\kappa_c$). As can
be seen from the partially-integrated results, even the lightest
charm-quark-mass lattice data used here completely saturates the
integral and therefore there is no need to perform any tail-extension
procedure, and just the lattice (trapezoid-rule) integral
suffices. 
We also note that the overall integrand and integral becomes substantially smaller as
$\kappa_c$ decreases, representing the fact that this integral must
vanish in the limit $\kappa_c\rightarrow 0$. There is a strong
negative tail in the integrand causing a fairly significant
cancellation for the overall integral, which becomes smaller as the
charm-mass decreases. At very low $\kappa_c$ on coarse lattices it is
unlikely that we will be able to properly resolve the peak of the
integrand and end up with a lower estimate due to the negative tail
cancelling against the peak contribution more than it should.  As we
move to finer lattices and the resolution at low $|y|$ improves, we
resolve the peak structure much better, as illustrated in
Fig.~\ref{fig:JBOX_cont}.

\subsection{Mass-dependence of the connected contribution}

\begin{figure}[h!]
\includegraphics[scale=0.5]{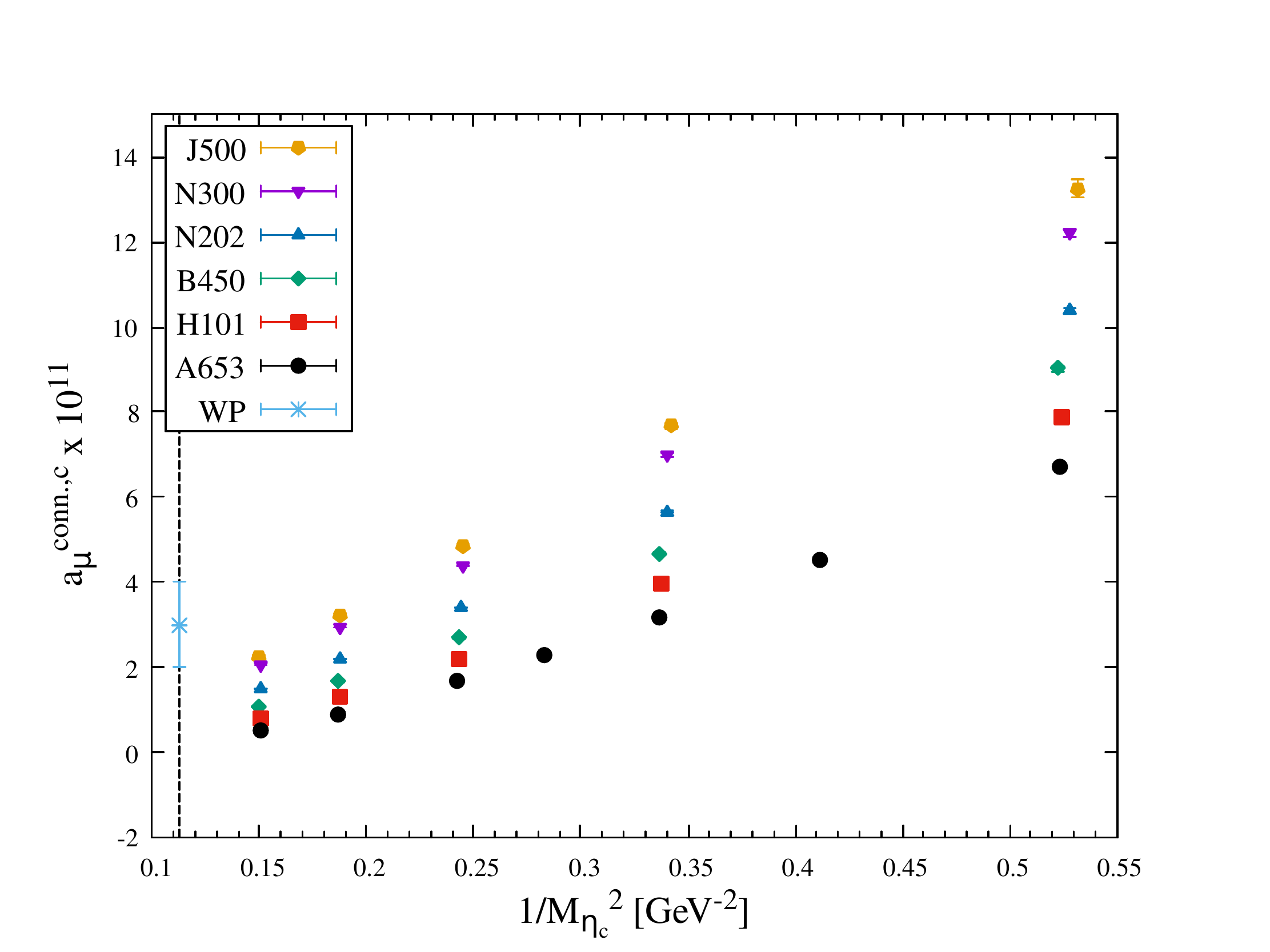}
\\\vspace{12pt}
\includegraphics[scale=0.5]{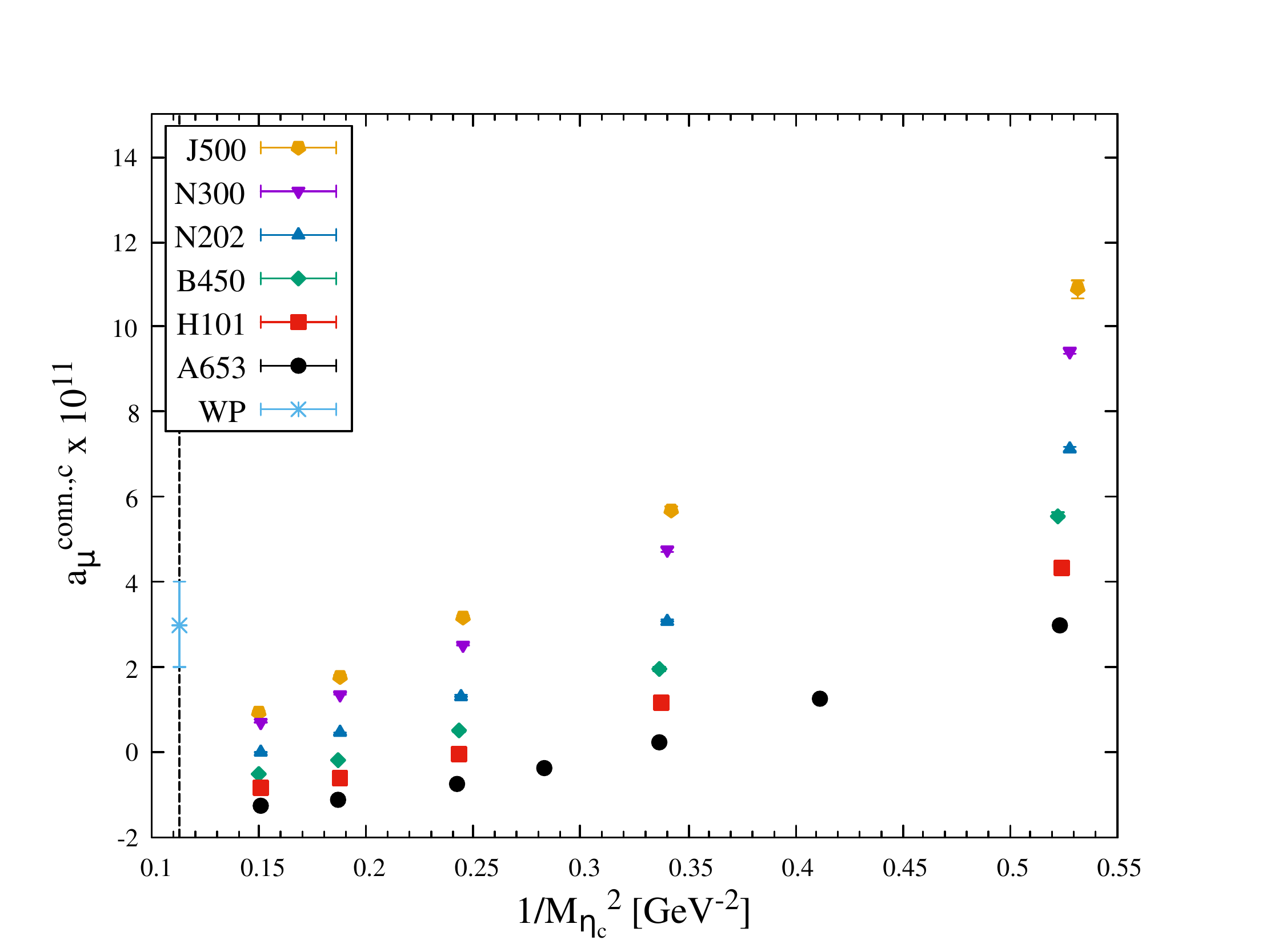}
\caption{Lattice results for the connected part of $\ahlbl$ in units of $10^{11}$ for the local-conserved (top) and local-local (bottom) data (see text). The black vertical line on the left indicates the physical value of $1/M_{\eta_c}^2$. The light blue point lying at physical $1/M_{\eta_c}^2$ is the estimate from Ref.~\cite{Aoyama:2020ynm}.}\label{fig:lat_res}
\end{figure}

The results are given in Tabs.~\ref{tab:res_A653}~--~\ref{tab:res_J500} of Appendix~\ref{sec:data_tables_conn}
and summarized in Fig.~\ref{fig:lat_res}.
Expectations are that $a_\mu$ scales with $m_\mu^2/m_{\rm{heavy}}^2$~\cite{Beresetskii:1956, Cowland:1958}, so it is instructive to focus on the dependence of $\ahlblc$ on $1/\metac^2$.
The data show a clear monotonic decrease as $1/\metac^2$ is decreased
toward its physical value, starting (for the lightest charm quarks) at
or above the WP prediction and ending (for the heaviest charm quarks)
at or below the WP value. At similar $\eta_c$ mass, the data have a
large spread between the coarsest and finest ensemble, indicating
strong discretisation effects.

At this point, it is useful to compare the two choices of
discretisations for the currents: the spread is larger in the
local-local data than in the local-conserved data. Furthermore, the
curvature in $1/\metac^2$ has a stronger dependence on the lattice
spacing in the local-local data. In addition, the fact that the coarse
local-local data at large $\metac$ become negative makes it more
difficult to describe the data using a fit ansatz.
For these reasons, we decide to base our determination of $\ahlblc$ solely on the analysis of the local-conserved data.

\subsection{Extrapolation to the continuum and to the physical charm mass}\label{sect:extrp}

Due to the heaviness of the valence charm quark, the intermediate
states that could potentially contribute to the correlation function
in question should be much suppressed at large distances;
see the discussion in section~\ref{sec:theo}.
Indeed, this can be seen by the saturation of the tail
of the lattice integrand (Figs.~\ref{fig:ABOX_cont} and~\ref{fig:JBOX_cont}).
For this reason, in the approach to the physical point, we assume that the
finite-size effects are minor and only extrapolate in the
$\eta_c$-meson mass and lattice spacing $a$.  The statistical error on
each individual data point is at the percent-level, which is
comparable to the quoted error on the lattice spacings given in
Tab.~\ref{tab:ensembles}; therefore, it is crucial to include the
error on the lattice spacing while performing an extrapolation to the
physical point.

To this end, a global fit is performed based on a Bayesian
approach~\cite{Lepage:2001ym}, where we promote each lattice spacing
to a fit-parameter and associate to it a Gaussian prior with the
central value and the width taken to be the quoted central value of
the lattice spacing $\bar{a}$ and its error $\Delta a$
respectively. Although the parameter space is small, constructing a
fit-ansatz with a $\chidof$ on the order of unity is in fact not an
easy task. After various attempts, we have identified two classes
of ans\"atze which are able to describe the data with reasonably good $\chidof$.

The most restrictive constraint that we deem important to fulfill is the $m_\mu^2/m_{\rm{heavy}}^2$ scaling of $\ahlblc$ in the presence of a heavy scale~\cite{Beresetskii:1956, Cowland:1958}.
It is natural to first consider the $\eta_c$-meson mass for such a scale.
A challenging part of the construction of fit-ans\"atze is to handle the apparent non-linear behavior in $1/\metac^2$ of the data (see Fig.~\ref{fig:lat_res}), which gradually gets milder as we go down in the lattice spacing.
This motivates our first class of ans\"atze, the \textit{P-class}, which consist of linear combinations of a leading term in $1/\metac^2$ and terms in $a^n f(\metac)$ with $n\in\mathbb{N}^*$ and $f$ an elementary function, treating the non-linearity in $1/\metac^2$ as a lattice artifact.

\begin{figure}[h!]
\includegraphics[scale=0.35]{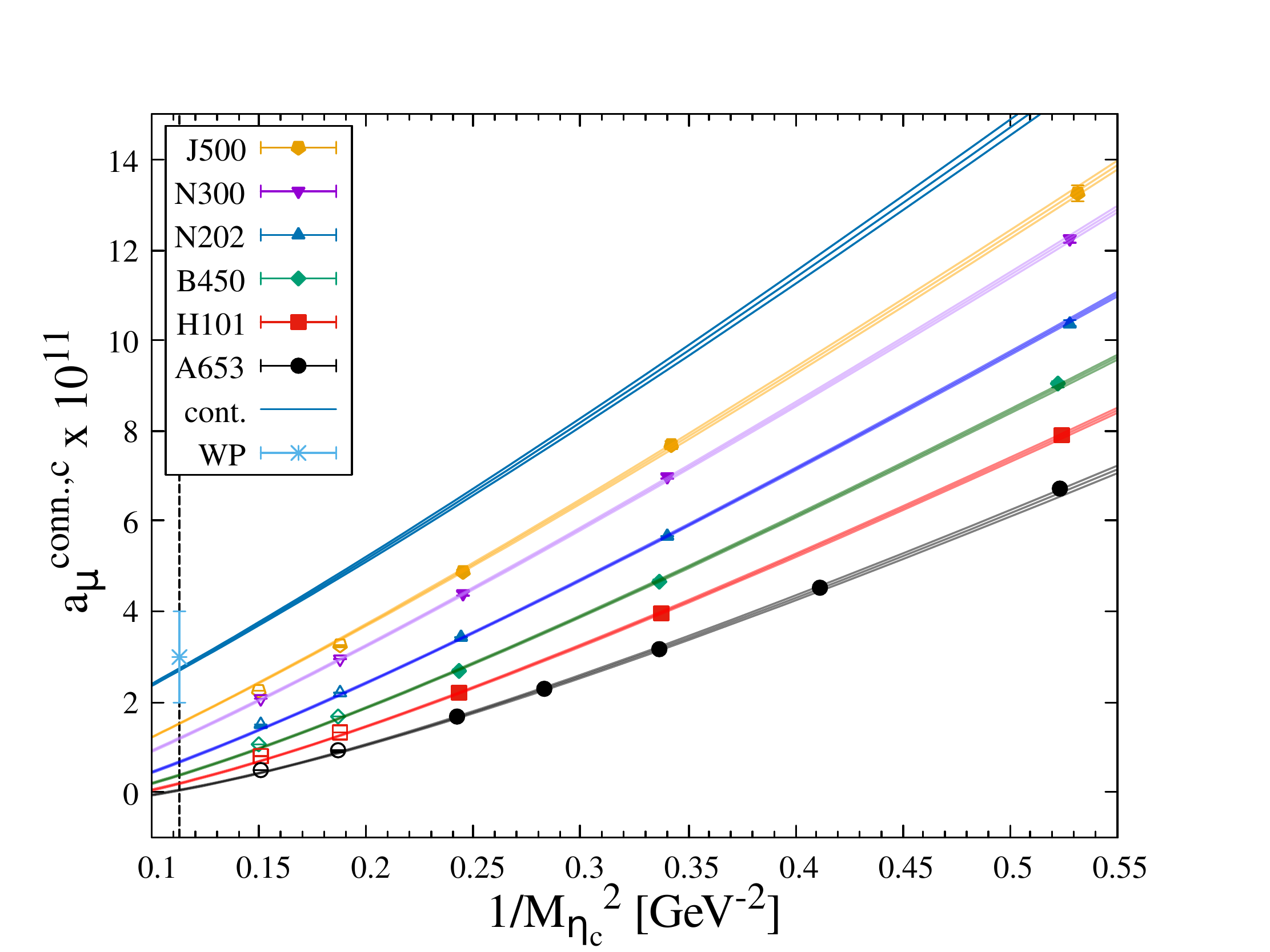}
\includegraphics[scale=0.35]{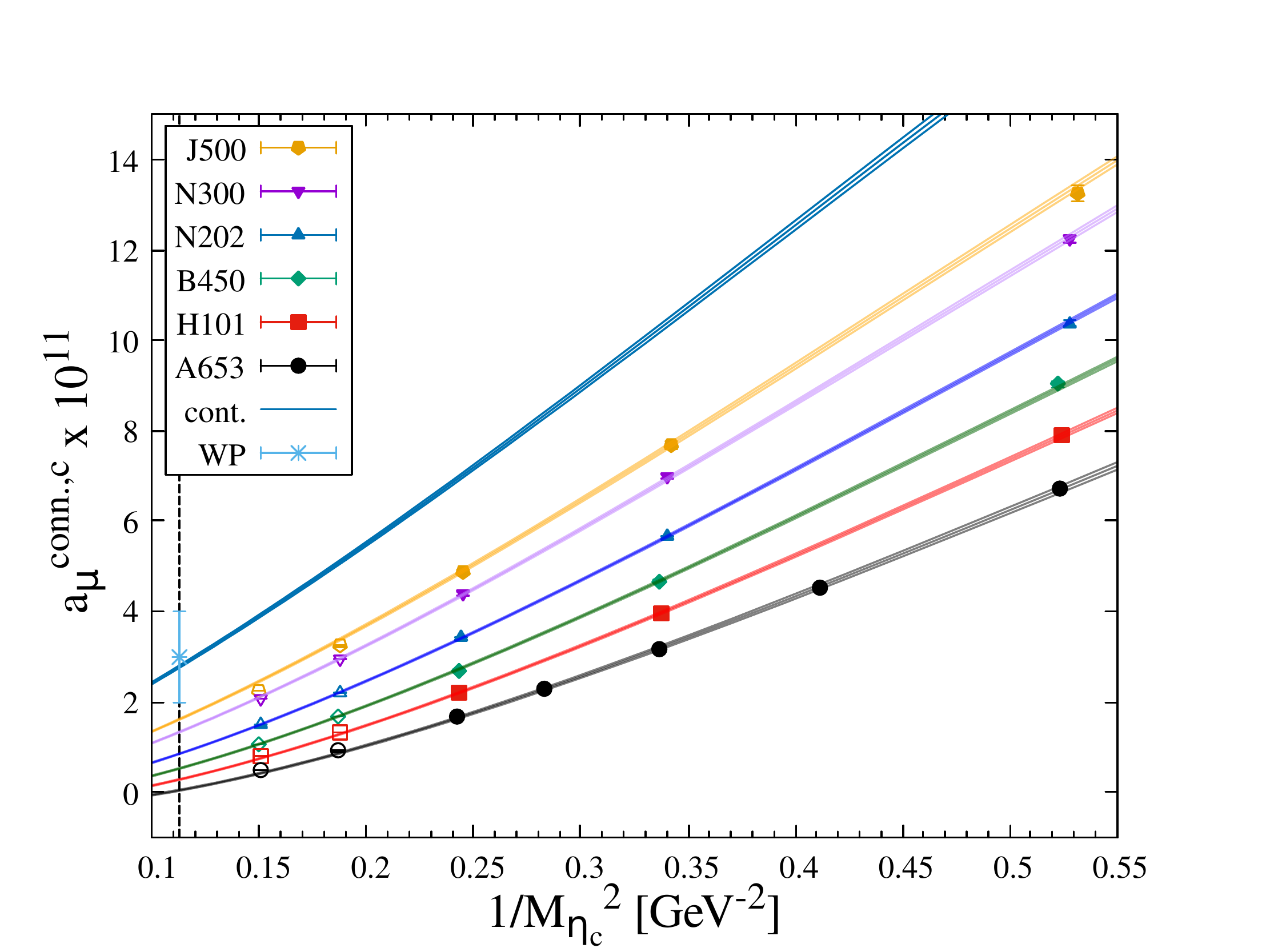}
\vspace{8pt}
\includegraphics[scale=0.35]{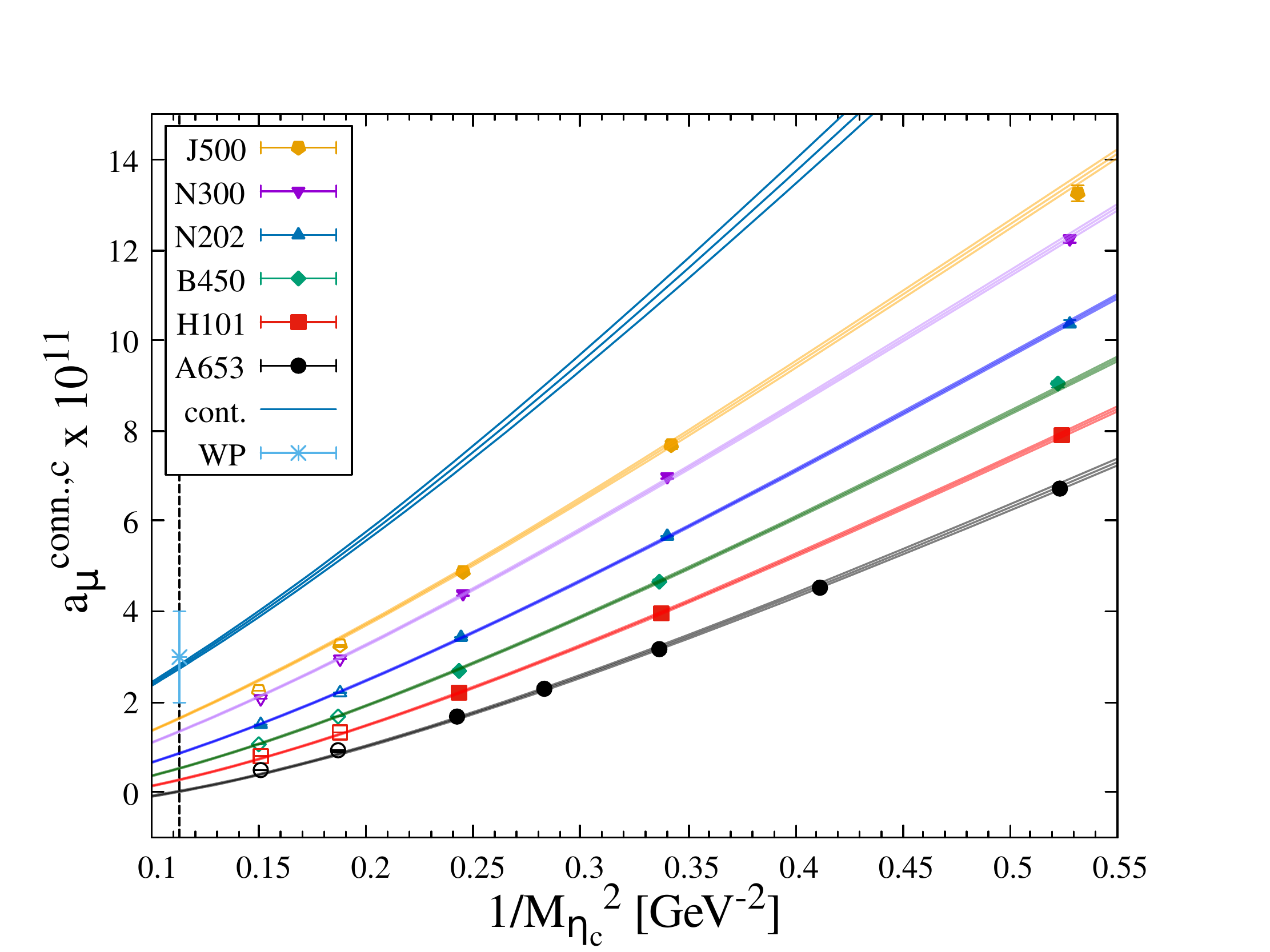}
\vspace{8pt}
\includegraphics[scale=0.35]{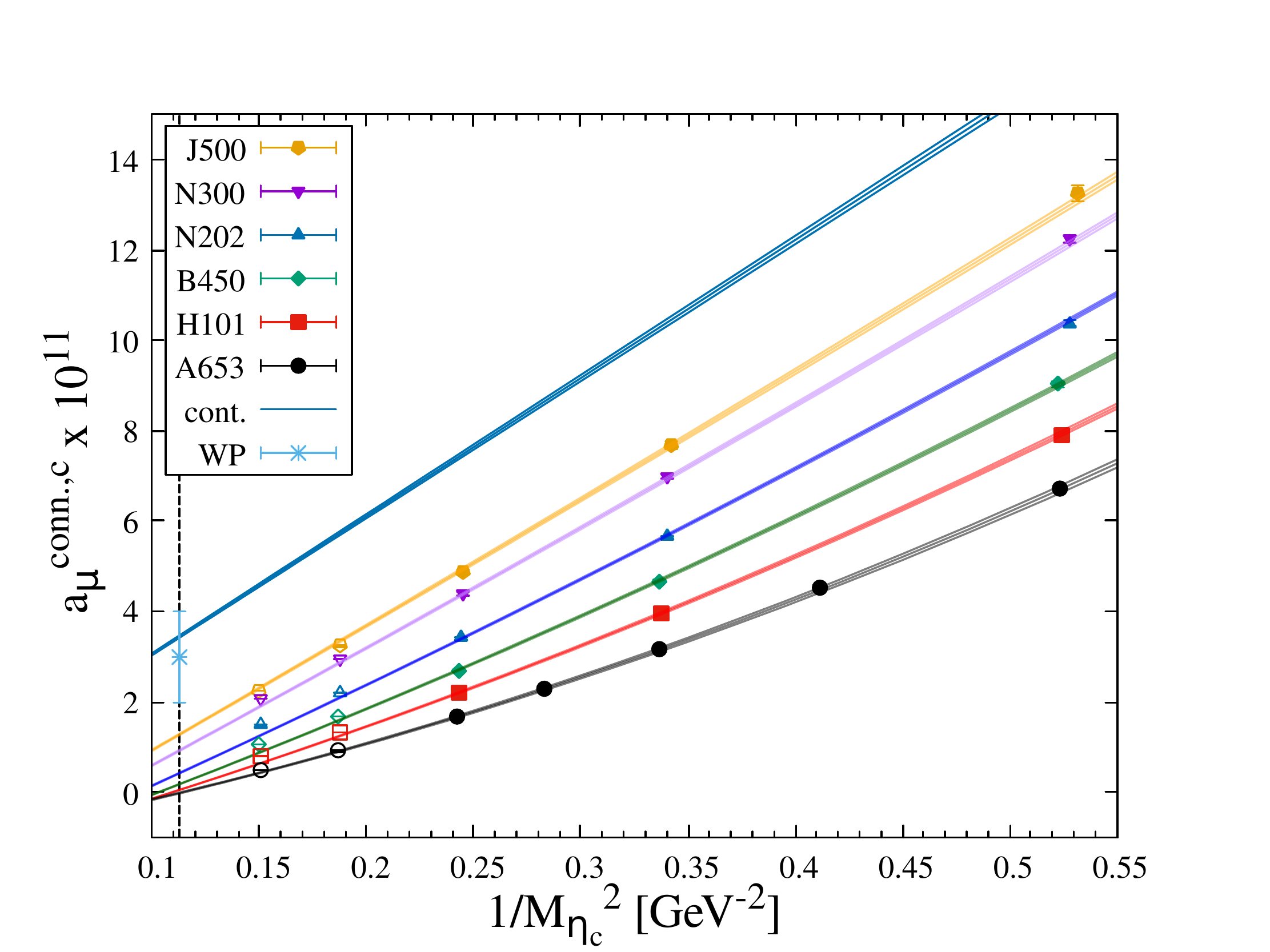}
\includegraphics[scale=0.35]{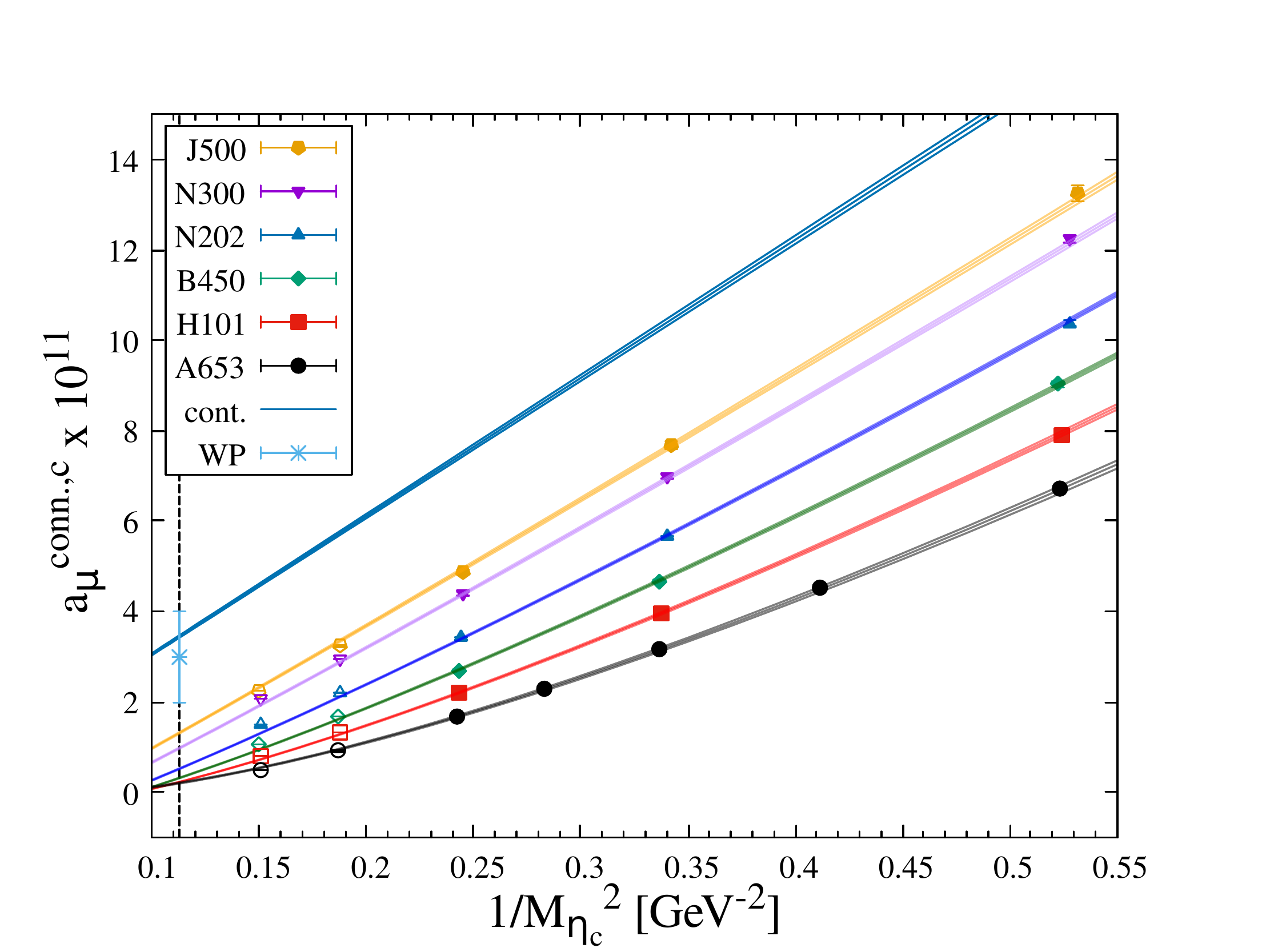}
\caption{Selected fit results with the dataset D1 (datapoints indicated by filled symbols): Fit 1, Fit 2, Fit 3, Fit 4, and Fit 5 (left to right from the top to the bottom). The continuum limit result is given by the top-most curve (``cont."). For other information, see Fig.~\ref{fig:lat_res}.}\label{fig:fit_res}
\end{figure}

Another way to account for the $m_\mu^2/m_{\rm{heavy}}^2$ scaling is to use the charm quark mass as a heavy scale. 
A rough estimate in the non-relativistic limit is that the $\eta_c$-meson mass should be equal to twice the charm quark mass, up to small relative corrections.
Based on this observation, we define the \textit{R-class} of fit ans\"atze, consisting of rational functions:
\begin{equation}\la{eq:Rclass}
\frac{P(a,\metac)}{\frac{1}{4}(C+\metac)^2 + Q(a,\metac)},
\end{equation}
where $P$ and $Q$ are polynomials in both $a$ and $\metac$ and $C$ is a constant.
In principle, $C$ can also have non-trivial dependence on $a$ and on $\metac$;
however, introducing additional parameters to describe this dependence turns out to be unnecessary,
as the non-linearity of the data can already be well captured with the form in Eq.\ (\ref{eq:Rclass}).

With the aforementioned two fit-ansatz classes, it remains nevertheless difficult to get reasonable $\chidof$ with the whole available dataset. 
In fact, this is not very surprising, as the resolution of the peak of the integrand becomes worse as $\kappa_c$ and $a$ become small (see Fig.~\ref{fig:ABOX_cont}).  
Therefore, it is necessary to allow for various cuts to the data.
At the same time, as we would like to reach as heavy as possible $\metac$ masses
in order to have a better control over the extrapolation to its physical value, it is preferable to discard as few data points as possible.
A lattice study in the pure QED case presented in App.~\ref{sec:lloop_app} shows that our setup should be valid up to a charm-quark mass of at least $20/3$ times that of the muon, with well-controlled cut-off effects.
Based on the latter and with a simple linear relation between the physical $\eta_c$ mass and the $\overline{\rm{MS}}$-mass of the charm quark~\cite{Zyla:2020zbs}, we demand an admissible fit to be able to cover the data points in the range of $1/\metac^2>0.20$ GeV${}^{-2}$.


Our fitting strategy goes as follows:
We build fit ans\"atze from either the P- or the R-class as explained earlier.
To avoid overfitting, the number of fit parameters is limited to five.
Apart from terms in $a^n M^m$, we have also tried logarithmic terms in $a$ or $M$ in order to allow for different types of curvature. The four datasets we consider are (D2, D3, and D4 are defined as D1 with extra omissions):
\begin{itemize}
\item D1: All ensembles and data points where $1/\metac^2>0.20 \text{ GeV}^{-2}$
\item D2: D1 where data from the coarsest lattice (A653) are omitted
\item D3: D1 where the lightest $\metac$ data of each ensemble are omitted
\item D4: D1 where $a\metac > 0.8$ are omitted 
\end{itemize}
Given that our final uncertainty estimate is dominated by systematics
due to the choice of fit ansatz and that attempts at correlated fits
yielded a poor fit quality, we choose to neglect correlations between
different $\metac$ on the same ensemble. Although this harms the
statistical interpretation of $\chi^2$ and $p$-values computed in the
standard way, we nevertheless use these to judge relative fit
quality. Our criterion for an admissible fit is one with a $p$-value
between 0.05 and 0.95, for which the extrapolated $a_\mu$ and the
$p$-value are stable under variation of the dataset choice.

We have tested various fit ans\"atze from both the P- and R-classes and found that the following five-parameter fits are able to describe our data with the quality requirements fulfilled (see Tab.~\ref{tab:res_fits} in Appendix~\ref{sec:fit_results} and Fig.~\ref{fig:fit_res}):
\begin{equation}\label{eq:fits}
\begin{aligned}
\textrm{Fit 1:}\quad\quad a_\mu(a,\metac) &= \frac{A+Ba\metac^2}{\frac{1}{4}(C+\metac)^2 + (D+Ea^2\metac^2)^2},\\
\textrm{Fit 2:}\quad\quad a_\mu(a,\metac) &= \frac{A+Ba\metac}{\frac{1}{4}(C+\metac)^2+(D+Ea^2\metac^2)^2},\\
\textrm{Fit 3:}\quad\quad a_\mu(a,\metac) &= \frac{A+Ba^2\metac^2}{\frac{1}{4}(C+\metac)^2+(D+Ea\metac^2)^2},\\
\textrm{Fit 4:}\quad\quad a_\mu(a,\metac) &= Aa + \frac{B+Ca^2}{\metac^2} + Da^2 + E\frac{a^2}{\metac^4},\\
\textrm{Fit 5:}\quad\quad a_\mu(a,\metac) &= Aa + \frac{B+Ca^2}{\metac^2} + Da^2 + E\frac{a^2}{\metac^2}\ln\metac.
\end{aligned}
\end{equation}
A further important feature of these fits is that they qualitatively follow the
trend of the data even in the region $1/\metac^2 < 0.2$ GeV${}^{-2}$.

As a general feature, the P-class ans\"atze tend to lead to larger results for $a_\mu(0,\metac^{\text{Phys}})$ as compared to the R-class. As there is no exclusive theoretical argument for the finite-lattice-spacing behaviour of these functions, and our data seem not to be able to unambiguously rule out any of these classes, our decision is to include the fit results with good $\chidof$ from both of them (see Tab.~\ref{tab:res_fits}). More specifically, our final result is the average of our lowest (Fit 1, D3 : 2.64(4)) and our largest (Fit 5, D2 : 3.47(3)) values and we assign a generous systematic error estimate by quoting half the difference of the two, which brings us to our estimate for the connected contribution,
\begin{equation}\label{eq:conn_res}
\ahlblc{}^{\text{,conn.}} = 3.1(4)\times 10^{-11}.
\end{equation}
Our error on this quantity is entirely dominated by the systematic error from our 
modeling of its dependence on $a$ and $M_{\eta_c}$.

\subsection{Comparison to the QED-based prediction}

\begin{figure}
\includegraphics[scale=0.55]{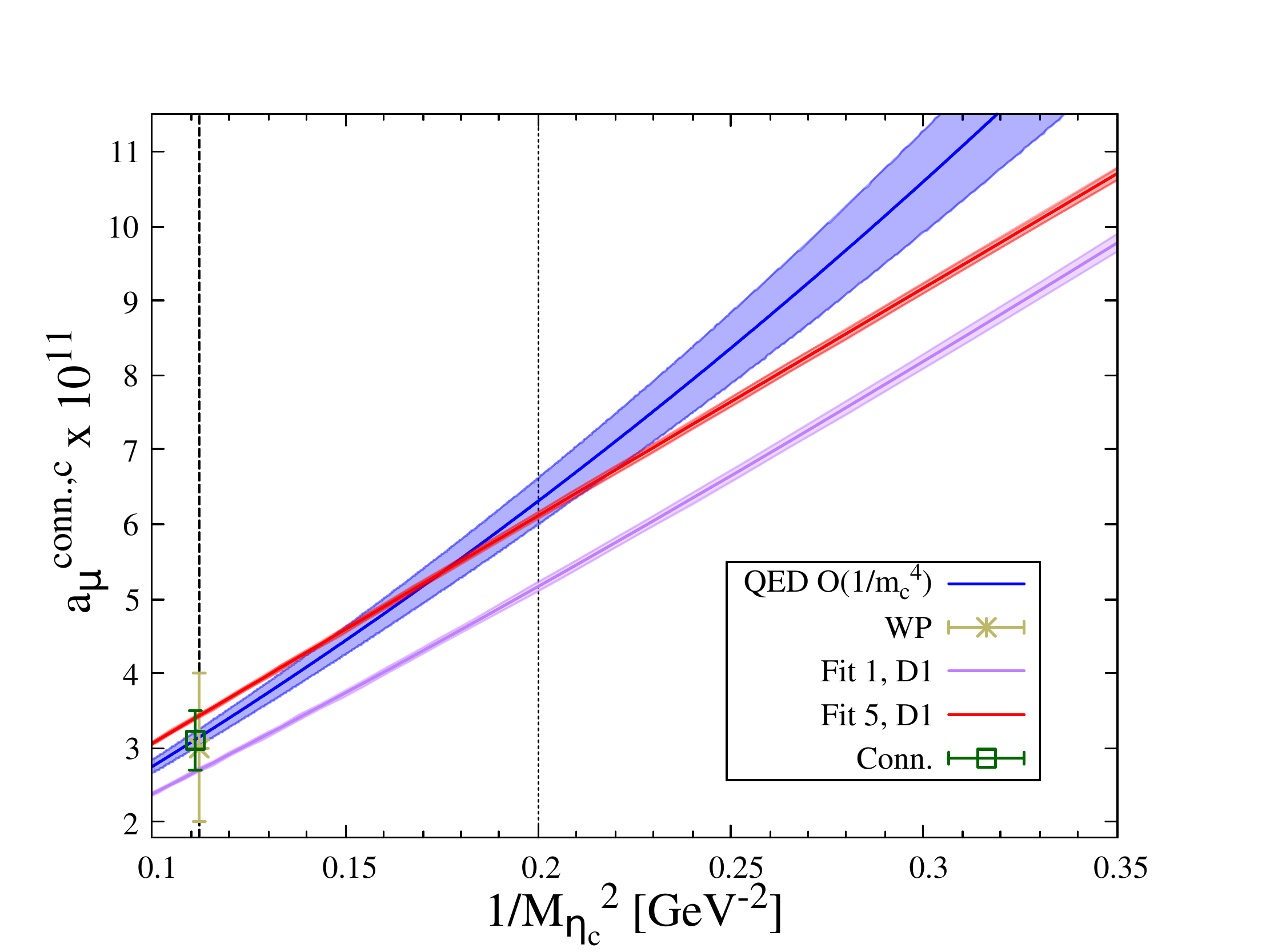}
\caption{The $\eta_c$-mass-dependence of $a_\mu^{\rm conn.,c}$: comparison between our continuum-extrapolated results using Fit~1 (lower, in magenta) and Fit~5 (higher, in red) based on dataset D1 and the QED-based prediction (in blue). The band on the latter indicates a $\pm5\%$ change in the mass-shift relating the charm-quark mass to the $\eta_c$ mass (see text). The dotted vertical line indicates the upper bound for the $\eta_c$ mass included in the fits. Our estimate for the connected contribution, Eq.~\eqref{eq:conn_res}, is marked with `Conn.' (a horizontal offset is applied for visibility).}
\label{fig:comp_qed}
\end{figure}

To close the study of the connected contribution, we compare our
result for $\ahlblc{}^{\text{,conn.}}$ to the charm-quark loop  evaluated analytically within QED
as given in Eq.\ (\ref{eq:leploop_analytic}).
To make contact with that expression, we need to specify the relationship between
the $\eta_c$ mass and the charm-quark mass.
As explained while discussing the R-class of
fit-ans\"atze, we assume the $\eta_c$ mass to be twice the charm quark
mass plus an almost charm-quark-mass independent mass-shift within a
given window of $\metac$.  We estimate the
mass-shift using the $\overline{\textrm{MS}}$ charm-quark mass and the
physical $\metac$ and assign a five-percent uncertainty to this
quantity.  
The prediction from this prescription is displayed in Fig.~\ref{fig:comp_qed}
together with our fit results.  
The difference
between truncating at O$(1/m_c^2)$ and at O$(1/m_c^4)$ is tiny
compared to the uncertainty that we assign to the mass-shift inferred
from our prescription.  It is worth noting that, even though the
QED-based prediction gives a result that falls in the bulk of our
estimate Eq.~\eqref{eq:conn_res} at the physical charm mass, the
milder curvatures in $1/\metac^2$ of the representative fit results
suggest that non-perturbative effects are still significant at intermediate masses.

\section{The disconnected contribution\la{sec:disc}}

The disconnected parts of the charm contribution are expected to be
very small.  From the outset, we neglect the (3+1),(2+1+1), and
(1+1+1+1) Wick-contraction topologies, based partly on them being
consistent with zero for the light quark contribution, as found
in~\cite{Chao:2021tvp}, and partly on the arguments laid out in
section \ref{sec:theo}.  This leaves us with the (2+2) topology, which
is a sizeable contribution in the light-quark $\ahlbl$ result. This
contribution can be broken into the mixed charm--light and the
charm--charm contributions, with the former (by analogy with the
strange sector) expected to be the major contribution.

As the disconnected contribution is still an expensive calculation, we
have limited ourselves to a single charm-quark mass determined by $\kappa_c$
from the $D_s$-tuning of Ref.\ \cite{gerardin:2019rua}.
This tuning is suboptimal for our present purposes,
an aspect we return to below. We will also use
the $Z_V^{c}$ values from~\cite{gerardin:2019rua}, except for ensemble A653, where we computed
the renormalisation factor ourselves. We employ exclusively local vector currents
and restrict ourselves to the ensembles N300, N202, B450, and A653, reusing data
for the light-quark loop from Ref.\ \cite{Chao:2020kwq}.

\begin{figure}[h!]
\includegraphics[scale=0.35]{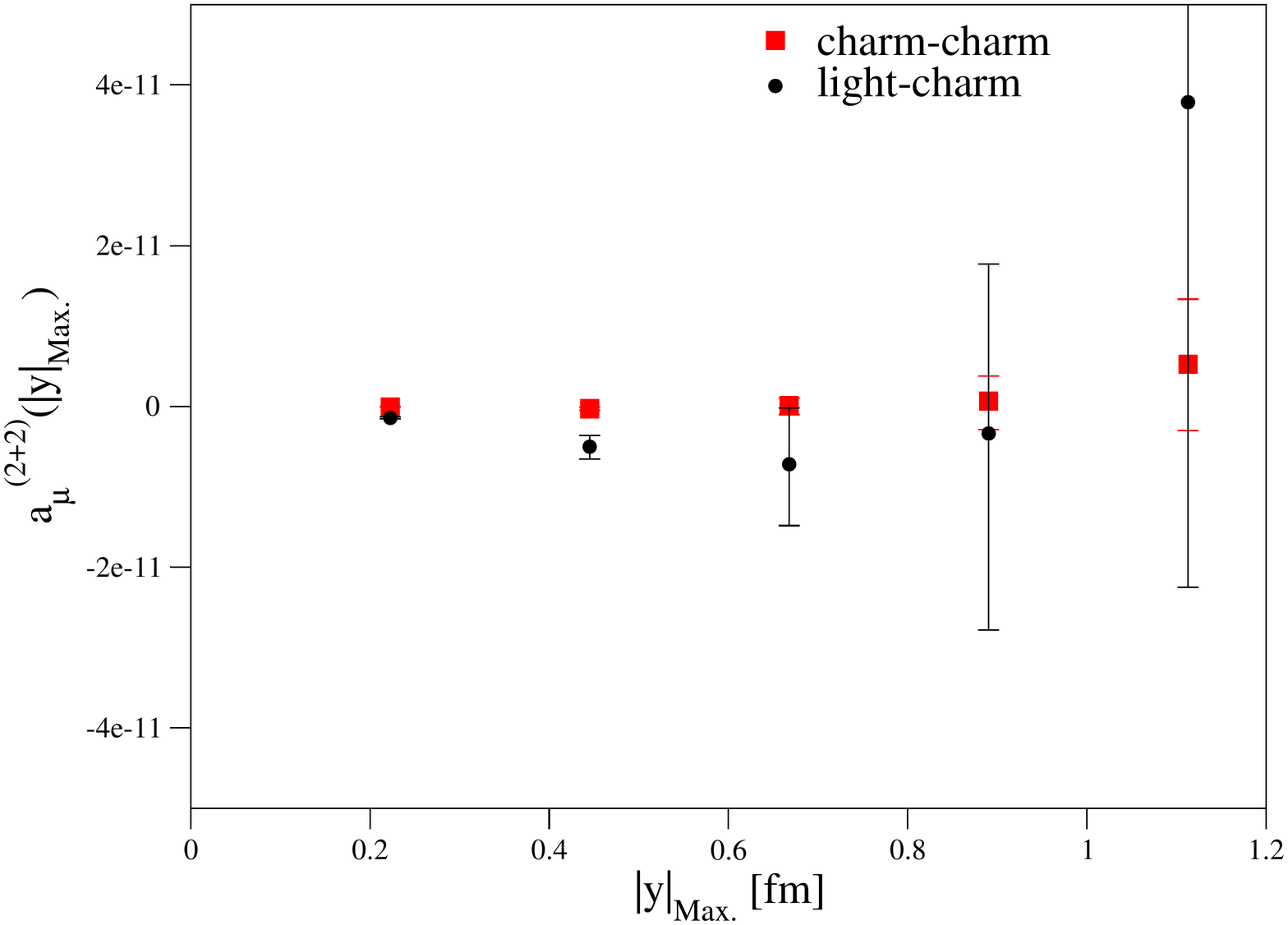}
\caption{The (2+2) charm--light and charm--charm partially-integrated data for ensemble N202.}\label{fig:N202_disc}
\end{figure}

A plot of the partially-integrated charm--light and charm--charm
disconnected contributions for ensemble N202 is shown in
Fig.~\ref{fig:N202_disc}. It is clear that both of these contributions
are noisy, small, negative, and very short-distance. Again this means
we can  use the lattice integral directly for our final result, and we use
a simple constant fit to the partially-integrated result for our final
determination. Based on the numerical evidence from Figs.\ \ref{fig:ABOX_cont}--\ref{fig:JBOX_cont}
that the connected contribution becomes very short-ranged as the charm mass is increased
towards its physical value, as well as the theoretical arguments of section \ref{sec:theo}, we start this fit
between 0.4 and 0.5\,fm. Tab.~\ref{tab:disc} shows our results for this
procedure and we see that A653 is an extreme outlier in the
charm--light contribution. The other, finer, ensembles yield values much smaller
and consistent with one another. We decide to omit this coarse
ensemble entirely and fit the remaining charm--light data to a straight
line in the variable $a^2$. This leads to the result
$a_\mu^{\text{HLbL,c,(2+2)}} = -0.28(21)\times 10^{-11}$.

We now come back to the issue of the tuning of the charm quark mass.
The CLS ensembles we are using are designed to have the trace of the quark mass matrix
equal to its physical value, to a rather good approximation~\cite{Bruno:2016plf}.
We remind the reader that for the connected contribution,  we chose the  charm-quark mass
such that the physical $\eta_c$ mass is reproduced. With this choice, the dependence of
charm correlators on the SU(3)$_{\rm f}$ breaking parameter $[m_s-(m_u+m_d)/2]$ is
expected to be small, being a pure sea quark effect. As a consequence, the extrapolation
to physical $(u,d,s)$ quark masses is expected to be very mild.
This is not the case if we tune the mass $\bar M_D$ of the triplet of $D$ mesons at our SU(3)$_{\rm f}$-symmetric point
to the physical $D_s$ meson, $M_{D_s}=1.968$\,GeV. By contrast, if we tune $\bar M_D$ to the average
$(\bar M_D)_{\rm phys} \equiv \frac{1}{3}[M_{D^+}+M_{D^0}+M_{D_s}]_{\rm phys}=1.901$\,GeV
of the physical $D$ meson masses, then we again avoid a valence-quark effect
in the approach to physical quark masses. It is also interesting to ask, how different a tuning this represents
as compared to the tuning via the $\eta_c$ mesons mass. We have found that the $\eta_c$ meson mass,
 extrapolated to the charm mass where  $\bar M_D= (\bar M_D)_{\rm phys}$,  amounts to $2.97(4)$\,GeV, which is consistent
with its physical value. This is an indication that sea quark effects are indeed small in the charm sector.

These observations lead us to apply a small correction to the charm--light disconnected contribution,
to bring it to the point where $\bar M_D$ takes the value $(\bar M_D)_{\rm phys}$.
Assuming that the disconnected contribution is roughly proportional to $1/\bar M_D^2$, we 
multiply our continuum-extrapolated result obtained at $\bar M_D = M_{D_s}$
with the ratio $(M_{D_s}/\bar M_D)_{\rm phys}^2$, leading to the final result
\be\label{eq:2p2_res}
a_\mu^{\text{HLbL,c,(2+2)}} = -0.30(23)\times 10^{-11}.
\ee
We neglect the charm--charm contribution as its contribution is far smaller than our final error for the charm--light.

\begin{table}[t]
\begin{tabular}{ccc|c|c}
\toprule
Ensemble & $\kappa_c$ & $Z_V^c$ & $a_\mu^{2+2:lc}\times 10^{11}$ & $a_\mu^{2+2:cc}\times 10^{11}$ \\
\hline
A653 & 0.119759 & 1.32265 & -3.24(99) & -0.06(2) \\
B450 & 0.125095 & 1.12972 & -0.53(27) & +0.01(2) \\
N202 & 0.127579 & 1.04843 & -0.48(14) & -0.03(2) \\
N300 & 0.130099 & 0.97722 & -0.39(8)  & -0.03(1) \\
\botrule
\end{tabular}
\caption{The charm--light and charm--charm (2+2) contributions tp $\ahlbl$.}\label{tab:disc}
\end{table}

\section{Discussion of the results and conclusion\la{sec:concl}}


We have determined the charm-quark contribution to hadronic
light-by-light scattering in the anomalous magnetic moment of the
muon. We find that the lattice determination of this quantity is
challenging, specifically in the modeling of the connected
contribution's discretisation effects: the associated systematic error
dominates our final error budget. As expected from the charm-loop
picture, the connected contribution turns out to be the most
significant overall.  We find the charm--light disconnected
contribution to be negative and much smaller in magnitude than the
fully-connected contribution, amounting to a $10\%$ correction with a
large uncertainty. The charm--charm disconnected contribution is
entirely negligible and we expect all higher-order contributions to be
equally insignificant.

Before quoting our final result for the charm contribution to
$\ahlbl$, we address the question of its dependence on the $(u,d,s)$
quark masses. The fact that several aspects of our lattice results can
be understood via the the charm-quark loop picture is an indication
that this dependence must be modest, and we may attempt to estimate
its order of magnitude via the ambiguity induced by the choice of the
charm-quark tuning condition away from the physical $(u,d,s)$
quark-mass point.  We saw in section \ref{sec:disc} that tuning the
average $D^+$, $D^0$ and $D_s$ mass to its physical value was
equivalent, within our uncertainties, to tuning the $\eta_c$ mass to
its physical value. Still, we estimate that the connected contribution
would potentially be modified by 2\% had we chosen the alternative
tuning.  Another estimate can be based on the idea that the charm
contribution is proportional to the sum of the inverses of the charged
$D$-meson squared masses.  That sum differs again by about two percent
from the square inverse of the average $D^+$, $D^0$ and $D_s$
mass. This argument suggests an absolute systematic error of
$0.06\times 10^{-11}$, which we conservatively inflate to $0.12\times
10^{-11}$ and add in quadrature to the other uncertainties below.

Our full result from adding Eqs.~\eqref{eq:conn_res} and
\eqref{eq:2p2_res} together and adding errors in quadrature is
\begin{equation}\la{eq:finalres}
\ahlblc = (2.8\pm 0.5)\times 10^{-11}.
\end{equation}
This result is completely consistent with the 2020 White Paper 
estimate of $(3\pm 1)\times 10^{-11}$~\cite{Aoyama:2020ynm}, and has half its uncertainty.

Combining Eq.\ (\ref{eq:finalres}) with our previous result from the light and strange
contributions of $a_\mu^{\text{HLbL,ls}}=(106.8\pm 15.9)\times 10^{-11}$
\cite{Chao:2021tvp} obtained with dynamical $(u,d,s)$ quarks yields a
fully non-perturbative determination of $\ahlbl$,
including all relevant contributions.
A last effect not yet accounted for is the charm sea-quark effect on the light-quark contributions,
as for instance the $D^+$ meson loop can contribute to the connected four-point function of the
down quark in a calculation with dynamical $(u,d,s,c)$ quarks. Within a scalar-QED treatment of the $D$ meson,
we have however estimated this effect to be below $0.1\times 10^{-11}$.
Therefore we neglect the charm sea quark effects and arrive at
\begin{equation}
\ahlbl=(109.6\pm 15.9)\times 10^{-11}.
\end{equation}
This concludes our first-generation calculation of hadronic light-by-light scattering in the muon $(g-2)$.

\acknowledgments{We thank Andreas Nyffeler for a fruitful collaboration on computing the QED kernel
  used here.  This work is supported by the European Research Council (ERC) under the
  European Union's Horizon 2020 research and innovation programme
  through grant agreement 771971-SIMDAMA, and through
 the Cluster of Excellence \emph{Precision Physics, Fundamental Interactions, and Structure of Matter} (PRISMA+ EXC 2118/1)  within the German Excellence Strategy (Project ID 39083149). The project leading to this publication has also received funding from the Excellence Initiative of Aix-Marseille University - A*MIDEX, a French “Investissements d’Avenir” programme, AMX-18-ACE-005.
  JRG acknowledges support from the Simons Foundation through the Simons Bridge for Postdoctoral Fellowships scheme.
 Calculations for this project were performed on the HPC clusters ``Clover'' and ``HIMster II'' at the Helmholtz-Institut Mainz and ``Mogon II'' at JGU Mainz.
Our programs use the deflated SAP+GCR solver from the openQCD package~\cite{Luscher:2012av}, as well as the QDP++ library
\cite{Edwards:2004sx}.
We are grateful to our colleagues in the CLS initiative for sharing ensembles.

 \appendix


\section{Methodology test for a heavy lepton}\label{sec:lloop_app}

With our implementation of the QED coordinate-space kernel, we have
been able to reproduce various known light-by-light contributions in
the continuum~\cite{Asmussen:2019act, Chao:2020kwq, Asmussen:2017bup}
at the one-percent level. The tests performed so far concern physics
involving particles with masses on the same order as the muon mass.
As our implementation of the QED-kernel relies on interpolating weight
functions that are precomputed on a grid~\cite{Asmussen:2016lse}, it
is important for the goal of this paper to test how robust this
implementation is for computing contributions from more massive
particles.

\begin{figure}[h!]
\includegraphics[width=0.70\textwidth]{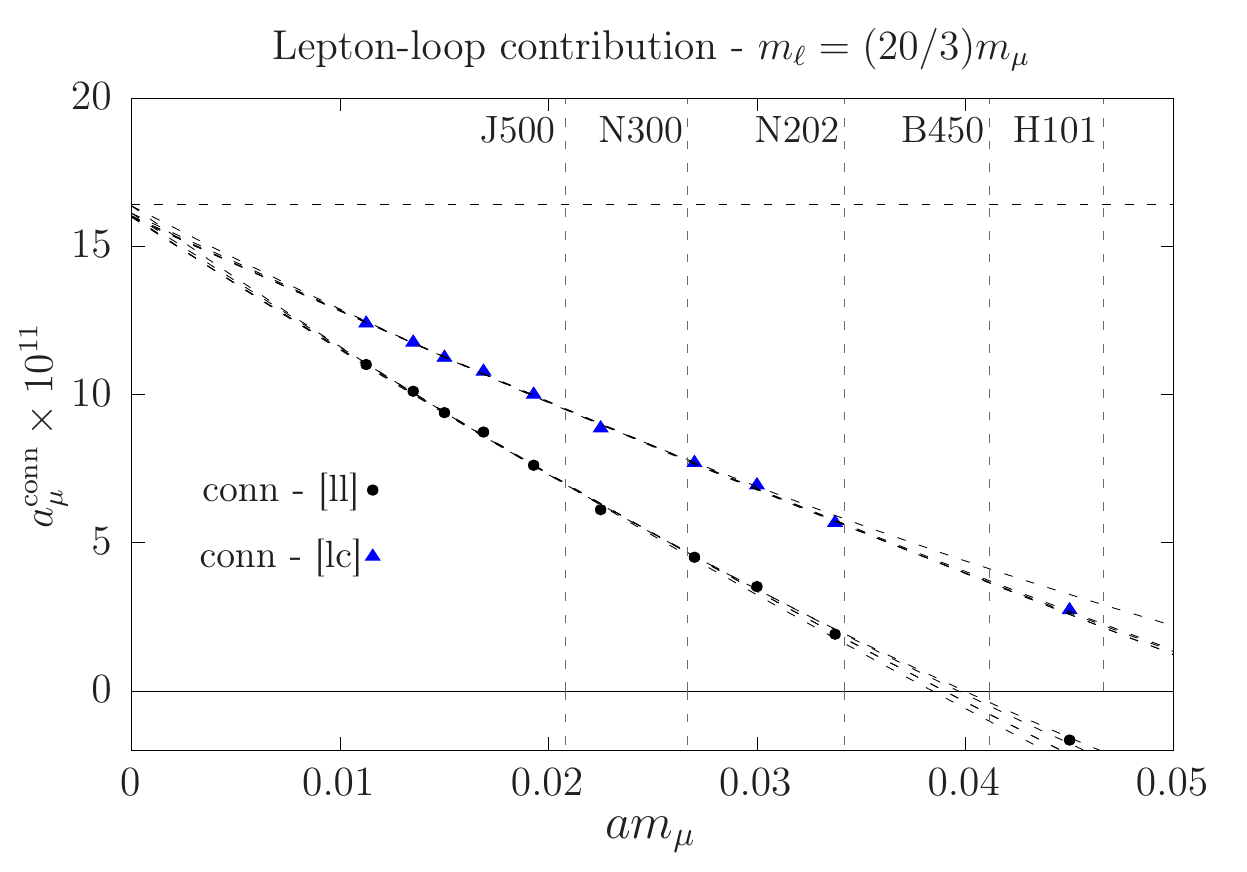}
\caption{Continuum extrapolation for the lepton loop with $m_\ell/m_\mu = 20/3$, computed on the lattice with two discretisations
of the vector current (local or conserved) at the external vertex $z$.}
\label{fig:leptloop20ov3}
\end{figure}

As an example of a calculation performed entirely in the continuum,
we quote the result we obtain with the kernel $\bar{\cal
L}^{(\Lambda=0.40)}$ and `method 2' for the lepton-loop contribution with
$m_\ell/m_\mu = 4$, namely $\ahlbl=(42.1\pm 0.5)\times 10^{-11}$; the
exact result is $(43.175...)\times 10^{-11}$.
While this precision is sufficient for our present purposes,
it is clear that, using continuum propagators for the lepton loop, the quality and stability of
the coordinate-space results degrade when the lepton
mass increases.\footnote{In our estimate, we regulate the numerics by setting the integrand to zero when
two vertices come within a distance of $10^{-3}m_\mu^{-1}$.
A more sophisticated procedure should be used for higher lepton masses.}

In order to validate our computational setup, we turn to a test that is much closer to
the procedure we used for the charm-quark contribution in lattice QCD.
We have computed the lepton-loop contribution to $a_\mu^{\text{HLbL}}$
using lattice fermion propagators at a mass-scale
relevant for this project, choosing specifically $m_\ell/m_\mu =
20/3$.  We proceed by repeating the calculation on increasingly fine
lattices and perform a continuum extrapolation using a quadratic
polynomial in $am_\mu$.  Here, two discretisations of the vector
current at the external vertex $z$ were used, and the resulting
contributions to $a_\mu^{\rm LbL}$ were extrapolated simultaneously to
the continuum, enforcing a common continuum value.
The deviation of the continuum-extrapolated result from the known exact result of $16.395..\times 10^{-11}$
depends somewhat on the choice of the extrapolation range, but is in all cases within 2.5\%.
This successfully passed test gives us confidence that the setup used for the lattice-QCD calculation presented  in the main text
is robust for fermion masses up to at least 700\,MeV.

\section{Tables of Data for the Connected Contribution}\label{sec:data_tables_conn}

This appendix contains the tables \ref{tab:res_A653} through \ref{tab:res_J500} providing
the results for the connected charm contribution to $\ahlbl$, using the 
`local-local' [ll]  and `local-conserved' [lc] discretisations.

\begin{table}[h!]
\begin{tabular}{cc|c|cc}
\toprule
$\kappa_c$ & $\metac$ [GeV] & $Z_V^{(c)}$ & $a_\mu^{\textrm{(Conn.[lc])}}$ & $a_\mu^{\textrm{(Conn.[ll])}}$ \\
\hline
0.1239619413500 & 2.5780 & 1.15128(7) & 0.494(9) & -1.246(26) \\
0.1260635511250 & 2.3144 & 1.06998(7) & 0.909(13) & -1.110(33) \\ 
0.1281651609000 & 2.0316 & 0.99145(6) & 1.657(23) & -0.726(42) \\
0.1292159657875 & 1.8815 & 0.95307(6) & 2.268(31) & -0.359(48) \\
0.1302667706750 & 1.7243 & 0.91530(6) & 3.163(44) & 0.238(58) \\
0.1313175755625 & 1.5589 & 0.87813(6) & 4.527(61) & 1.236(73) \\
0.1323683804500 & 1.3827 & 0.84154(9) & 6.717(106) & 2.973(101) \\
\botrule
\end{tabular}
\caption{Results for ensemble A653, $a_\mu$ has been multiplied by $10^{11}$.
\label{tab:res_A653}}
\end{table}

\begin{table}[h!]\label{tab:H101_conn}
\begin{tabular}{cc|c|cc}
\toprule
$\kappa_c$ & $\metac$ [GeV] & $Z_V^{(c)}$ & $a_\mu^{\textrm{(Conn.[lc])}}$ & $a_\mu^{\textrm{(Conn.[ll])}}$ \\
\hline
0.1263626550 & 2.5764 & 1.07159(7) & 0.811(7)  & -0.827(15) \\
0.1280954825 & 2.3112 & 1.00793(7) & 1.313(10) &  -0.588(19)\\
0.1298283100 & 2.0282 & 0.94610(6) & 2.205(16) &  -0.059(25)\\
0.1315611375 & 1.7212 & 0.88580(7) & 3.952(21) & 1.168(36)  \\
0.1332939650 & 1.3811 & 0.82708(6) & 7.901(69) & 4.328(70)  \\
\botrule
\end{tabular}
\caption{Same as Tab.~\ref{tab:res_A653} but for ensemble H101.}
\end{table}

\begin{table}[h!]\label{tab:B450_conn}
\begin{tabular}{cc|c|cc}
\toprule
$\kappa_c$ & $\metac$ [GeV] & $Z_V^{(c)}$ & $a_\mu^{\textrm{(Conn.[lc])}}$ & $a_\mu^{\textrm{(Conn.[ll])}}$ \\
\hline
0.128043750 & 2.5817 & 1.02163(4) & 1.073(7)  & -0.521(14) \\
0.129517875 & 2.3144 & 0.96948(3) & 1.663(10) &  -0.190(18)\\
0.130992500 & 2.0298 & 0.91842(3) & 2.696(17) &  0.488(22) \\
0.132466875 & 1.7228 & 0.86851(3) & 4.676(30) & 1.963(34)  \\
0.133941250 & 1.3830 & 0.81977(4) & 9.037(65) & 5.558(65)  \\
\botrule
\end{tabular}
\caption{Same as Tab.~\ref{tab:res_A653} but for ensemble B450.}
\end{table}

\begin{table}[h!]\label{tab:N202_conn}
\begin{tabular}{cc|c|cc}
\toprule
$\kappa_c$ & $\metac$ [GeV] & $Z_V^{(c)}$ & $a_\mu^{\textrm{(Conn.[lc])}}$ & $a_\mu^{\textrm{(Conn.[ll])}}$ \\
\hline
0.129934250 & 2.5763 & 9.68482(3) & 1.483(7)  & -0.021(10) \\
0.131111875 & 2.3077 & 0.92860(3) & 2.199(10) & 0.444(12) \\
0.132289500 & 2.0227 & 0.88955(6) & 3.409(16) & 1.322(16) \\
0.133467125 & 1.7154 & 0.85096(2) & 5.650(28) & 3.094(26) \\ 
0.134644750 & 1.3760 & 0.81322(2) & 10.381(55) & 7.115(55) \\
\botrule
\end{tabular}
\caption{Same as Tab.~\ref{tab:res_A653} but for ensemble N202.}
\end{table}

\begin{table}[h!]\label{tab:N300_conn}
\begin{tabular}{cc|c|cc}
\toprule
$\kappa_c$ & $\metac$ [GeV] & $Z_V^{(c)}$ & $a_\mu^{\textrm{(Conn.[lc])}}$ & $a_\mu^{\textrm{(Conn.[ll])}}$ \\
\hline
0.131824250 & 2.5779 & 0.92043(8) & 2.066(12) & 0.695(15) \\
0.132686875 & 2.3074 & 0.89253(4) & 2.943(17) & 1.361(19) \\
0.133549500 & 2.0209 & 0.86506(3) & 4.390(27) & 2.529(28) \\ 
0.134412125 & 1.7137 & 0.83794(3) & 6.980(47) & 4.733(44) \\
0.135274750 & 1.3763 & 0.81118(3) & 12.236(95)& 9.425(81) \\
\botrule
\end{tabular}
\caption{Same as Tab.~\ref{tab:res_A653} but for ensemble N300.}
\end{table}

\begin{table}[h!]\label{tab:J500_conn}
\begin{tabular}{cc|c|cc}
\toprule
$\kappa_c$ & $\metac$ [GeV] & $Z_V^{(c)}$ & $a_\mu^{\textrm{(Conn.[lc])}}$ & $a_\mu^{\textrm{(Conn.[ll])}}$ \\
\hline
0.132950861040 & 2.5849 & 0.89274(1) & 2.245(19) & 0.950(20) \\
0.133601050870 & 2.3090 & 0.87241(1) & 3.239(27) & 1.781(28) \\
0.134251240700 & 2.0194 & 0.85228(1) & 4.860(45) & 3.183(43) \\
0.134901430525 & 1.7102 & 0.83240(4) & 7.686(77) & 5.707(74) \\
0.135551620350 & 1.3718 & 0.81266(1) & 13.255(196) & 10.890(197) \\
\botrule
\end{tabular}
\caption{Same as Tab.~\ref{tab:res_A653} but for ensemble J500.}
\label{tab:res_J500}
\end{table}

\newpage

\section{Fit results for the connected contribution}\label{sec:fit_results}

In this appendix we collect details of the fit results to the local-conserved connected data
obtained with the fit ans\"atze of Eq.\ (\ref{eq:fits}). The values obtained for $\ahlblc{}^{\rm{,conn.}}$,
as well as the corresponding $\chidof$ and $p$ values are given in Table~\ref{tab:res_fits}.

\begin{table}[h!]
\begin{tabular}{cc|c|c|c}
\toprule
Fit & Dataset & $\ahlblc{}^{\rm{,conn.}}\times 10^{11}$ & $\chidof$ & $p$-value \\
\hline
Fit1 & D1  &  2.71(3) & 1.08 & 0.37 \\
Fit1 & D2  &  2.71(3) & 1.18 & 0.30 \\
Fit1 & D3  &  2.64(4) & 1.07 & 0.38 \\
Fit1 & D4  &  2.76(3) & 1.27 & 0.23 \\
\hline 
Fit2 & D1  &  2.77(3) & 1.11 & 0.34 \\
Fit2 & D2  &  2.81(3) & 1.37 & 0.19 \\
Fit2 & D3  &  2.72(3) & 1.14 & 0.33 \\
Fit2 & D4  &  2.85(3) & 1.03 & 0.42 \\
\hline
Fit3 & D1  &  2.77(5) & 1.59 & 0.07 \\
Fit3 & D2  &  2.72(5) & 1.98 & 0.03 \\
Fit3 & D3  &  2.84(4) & 1.74 & 0.07 \\
Fit3 & D4  &  2.66(5) & 1.18 & 0.30 \\
\hline
Fit4 & D1  &  3.43(3) & 1.51 & 0.09 \\
Fit4 & D2  &  3.47(3) & 1.48 & 0.14 \\
Fit4 & D3  &  3.43(3) & 1.90 & 0.05 \\
Fit4 & D4  &  3.45(3) & 1.97 & 0.03 \\
\hline
Fit5 & D1  &  3.43(3) & 1.48 & 0.10 \\
Fit5 & D2  &  3.47(3) & 1.48 & 0.14 \\
Fit5 & D3  &  3.42(3) & 1.92 & 0.04 \\
Fit5 & D4  &  3.45(3) & 1.94 & 0.03 \\
\botrule
\end{tabular}
\caption{Fit results of the local-conserved connected charm contribution to $a_\mu$.}
\label{tab:res_fits}
\end{table}

\bibliographystyle{apsrev4-1}
\nocite{bibtitles}
\bibliography{refs}

\end{document}